\documentclass[12pt]{article}

\def\hybrid{%
  \topmargin 0pt 
  \oddsidemargin 0pt \headheight 0pt \headsep 0pt
       \textwidth 6.5in        
       \textheight 9in         
  \marginparwidth 0.0in   %
  \parskip 5pt plus 1pt \jot = 1.5ex
}%
\hybrid

\usepackage[dvips]{graphicx} \usepackage{epsf}

\usepackage{varioref} %
\renewcommand{\vref}[1]{\ref{#1}~\vpageref{#1}\unskip}

\IfFileExists{srcltx.sty}{\usepackage[active]{srcltx}}

\usepackage{amsmath}
\usepackage{amsfonts}

\def\h{\hbar}
\newcommand{\la}{\langle}
\newcommand{\ra}{\rangle} 
\newcommand{\ket}[1]{\left|{#1}\right\rangle}
\newcommand{\bra}[1]{\left\langle{#1}\right|}

\newcommand{\sumk}{\sum\limits_{k=1}^\infty}
\newcommand{\pfrac}[2]{\displaystyle\frac{\p #1}{\p #2}}
\newcommand{\p}{\partial}
\renewcommand{\Re}{\mathop\mathrm{Re}}
\newcommand{\z}[1]{{z_{#1}}}
\newcommand{\bz}[1]{\bar z_{#1}}
\newcommand{\zb}[1]{\bar z_{#1}}

\newcommand{\ba}[1]{\ensuremath{{ \bar a}_{#1}}}

\newcommand{\s}[1]{s_{#1}}
\renewcommand{\sb}[1]{\bar s_{#1}}

\renewcommand{\t}[1]{t_{#1}}
\newcommand{\bt}[1]{\bar t_{#1}}
\newcommand{\tb}[1]{\bar t_{#1}}

\def\CC{\ensuremath{{\cal C}}}

\def\CE{{\cal E}}

\def\CJ{{\cal J}}

\def\CO{{\cal O}}

\def\bea{\begin{eqnarray}}
\def\eea{\end{eqnarray}}

\def \p{\partial}

\def \bz {{\bar z}}

\def \bw {{\bar w}}

\def\be{\begin{equation}}
\def\ee{\end{equation}}

\def \Tr {{\rm tr}}
\def\n{{N+1}}

\newcommand{\bket}[1]{\left|{#1}\rangle\rangle\right.}

\newcommand{\sprod}[2]{\langle{#1}| {#2}\rangle}
\newcommand{\res}{\mathop{\mathrm{res}}}

\newcommand{\0}{{(0)}}
 
\newcommand{\2}{{(2)}}

\newcommand{\trep}{$s$-representation}
\newcommand{\zrep}{$z$-representation{}}

\newcommand{\bigbar}[1]{\overline{\mathstrut #1}}

\catcode`\@=11
\def\numberbysection{\@addtoreset{equation}{section}
           \def\theequation{\thesection.\arabic{equation}}}
\catcode`\@=12

\numberbysection

\usepackage{hyperref}

\begin{document} 

\title{%
  \hfill\vbox{\normalsize\hbox{hep-th/0312242}%
    \hbox{AEI-2003-112}
    \hbox{EPFL-IPT-03}
    \hbox{IHES/P/03169}%
  }\\[12pt]
  \Large Classical and Quantum Branes in $c=1$ String Theory and Quantum Hall
  Effect}
\author{%
  Alexey Boyarsky$^a$\thanks{alexey.boyarsky@epfl.ch}~~ Bogdan
  Kulik$^b$\thanks{bogdan.kulik@aei.mpg.de}~~~and~
  Oleg Ruchayskiy$^c$\thanks{ruchay@ihes.fr}\\[12pt]
  {\normalsize \it $^a$ Laboratoire de physique des particules et
  cosmologie}\\
  {\normalsize \it Ecole Polytechnique F\'ed\'erale de Lausanne} \\
  {\normalsize \it BSP-Dorigny 1015, Lausanne, Switzerland}\\[12pt]
  {\normalsize \it $^b$Max-Planck-Institut f\"ur Gravitationsphysik}\\
  {\normalsize \it Albert-Einstein-Institut}\\
  {\normalsize \it Am M\"uhlenberg 1, D-14476 Golm, Germany}\\[12pt]
  {\normalsize $^c$\textit{Institut des Hautes Etudes Scientifiques}} \\
  {\normalsize \textit{Bures-sur-Yvette, F-91440, France}}
}%

\date{\footnotesize\today}
\maketitle

\begin{abstract}
  
  Interpretation of D1 and D0-branes in 1+1 string theory as classical and
  quantum eigen-values in dual c=1 Matrix Quantum Mechanics (MQM) was recently
  suggested. MQM is known to be equivalent to a system of $N$ free fermions
  (eigen-values).  By considering quantum mechanics of fermions in the
  presence of classical eigen-value we are able to calculate explicitly the
  perturbation of the shape of Fermi-sea due to the interaction with the
  brane.  We see that the shape of the Fermi-sea depending on the position of
  the classical eigen-value can exhibit critical behavior, such as development
  of cusp.  On quantum level we build explicitly the operator describing
  quantum eigen-value. This is a vertex operator in bosonic CFT. Its
  expectation value between vacuum and Dirichlet boundary state is equal to
  the correct wave-function of the fermion.  This supports the conjecture that
  quantum eigen-value corresponds to D0-brane.  We also show that $c=1$ MQM
  can be obtained as analytical continuation of the system of 2d electrons in
  magnetic field which is studied in Quantum Hall Effect.
\end{abstract}

\section{Introduction}
\label{Verl}

In the works \cite{Verl1,Verl2} an interesting interpretation of duality
between two-dimensional string theory and $c=1$ matrix model was suggested.

It was known for a long time that double-scaling limit (which in
particular implies large $N$ limit) of Matrix Quantum Mechanics (MQM)
is dual to two-dimensional string theory in linearly growing dilaton
background (see e.g.~\cite{klebanov-2d}). The matrix model
with finite $N$ was an auxiliary notion in this duality and did not
have any string-theoretic interpretation.

Proposal of \cite{Verl1,Verl2} is to identify the MQM of finite $N$
matrices with an effective action of open-string tachyon living
on D0 brane.  

As it is well-known the $c=1$ MQM is equivalent to the system
of $N$ free fermions moving in upside-down quadratic
potential. Its Hamiltonian is
\begin{eqnarray}
  \label{eq:1}
  H = \frac 12 \sum_{i=1}^N ({\hat p}_i^2 - {\hat x}_i^2 )
\end{eqnarray}
Eq.~(\ref{eq:1}) is a free system, thus all fermions occupy individual energy
levels up to some Fermi-level below the top of the potential.  The collective
excitations described by small deformations of the Fermi-level are dual to
closed string degrees of freedom in corresponding string theory
\cite{Polchinski}.  The new identification is a suggestion that fermions
sitting on top of the quadratic potential are dual to unstable D0 brane of the
string theory.  The important conceptual point is that the fermion dual to
D0-brane lies outside the Fermi-sea of the rest of electrons whose collective
excitations were earlier identified with closed strings.  The rolling-down of
the fermion from the top of the potential to Fermi-level thus describes the
decay of the brane.

This brings into play Sen's picture of decay of unstable branes
via tachyonic condensation. In accordance to Sen's conjectures
a brane decays into closed string vacuum radiating
closed strings in the process.

Let us discuss the MQM calculations of this process in some details.
The rolling down fermion should interact with fermions forming the
Fermi sea. In \cite{Verl1,Verl2} it is suggested 
to describe this interaction as a motion
of $N$ quantum fermions 
in external field of one classical fermion.
The introduction of the classical fermion should satisfy the following
consistency condition. One can calculate partition function
of the first $N$ fermions. It will depend on a coordinate of the external
classical fermion. If the path integral is taken with respect to it
as well one has to get back the partition function of $N+1$ free fermions
in upside-down oscillator potential.

To carry out this program let us consider partition function of $N$ free
fermions of MQM. (We set $\hbar=1$ in all formulas.  We will re-introduce it
later in the paper at the point of taking quasi-classical limit).
\begin{eqnarray}
Z_N(\tau) = \int \prod_{i=1}^N D\lambda_i(t) 
e^{i\int^\tau dt {\cal L}_N} \Delta_N(\tau) \label{part.N}
\end{eqnarray}
where $ {\cal L}_N$ is a Lagrangian of $N$ free fermions in the upside-down
oscillator potential. The presence of the Van der Monde determinant
$\Delta_N(\tau)$
\begin{equation}
\Delta_N(\tau)=\prod_{1\le i<j}^N(\lambda_i(\tau)-\lambda_j(\tau))\label{eq:2}
\end{equation}
at the end point
$\tau$ of integration changes the Hamiltonian of the system
to
\begin{eqnarray}
{\hat H} = \Delta_N H_N \Delta_N^{-1}~~~~~~
H_N = \sum_i (\frac14p_i^2 - \lambda_i^2)
\end{eqnarray}
The claim of \cite{Verl1,Verl2} is that $N$-quantum
fermions moving in the field of a classical one are described
by the similar Hamiltonian
\begin{eqnarray}
{\hat H}_{new} = \Delta_{N+1} H_N \Delta_{N+1}^{-1}
\end{eqnarray}
where the classical trajectory $\lambda_{N+1}(t)=z(t)$ sits only in the Van
der Monde $\Delta_{N+1}$. This new Hamiltonian has now an interaction between
fermions and can be written as
\begin{eqnarray}
\label{eq:18}
{\hat H}_{new} = \Delta_N( H_N + H_{int}) \Delta_N^{-1} \\
 \label{eq:42}
H_{int} = [H_N,\Phi(z(t))]\\
\label{eq:25}
\Phi(z(t)) = \log 
\left( 
\frac{\Delta_{N+1}(z)}{ \Delta_N} 
\right)
= \sum_i\log(\lambda_i - z)
\end{eqnarray}
Thus the interaction is just an exchange interaction between classical and
quantum fermions. Let us stress one more time that from the point of view of
the first $N$ free fermions it looks like an external field produced by decaying
``classical'' brane.  To get to the quantum picture one needs to perform path
integral over the classical trajectories $z(t)$ weighed by the classical
action.  Equivalently, one can introduce second-quantized local operator
creating a new particle. Comparing this first approach to eq.~(\ref{part.N})
one gets \cite{Verl2} for the second-quantized operator
\begin{eqnarray}
\Psi(z_0,t_0) = \int_{z(t_0)} {\cal D} z(t)
e^{i \int_{t_0}({\dot z}^2 + z^2 + H_{int}(z))} =
\psi(z_0,t_0) \exp(i {\hat \Phi}(z_0,t_0)) \label{Ver.bos}
\end{eqnarray}
where $\psi(z_0,t_0)$ is a wave function of a single free fermion.  Thus in
quantum treatment the operator $\Phi$ that defined interaction with one
classical fermion bosonizes the operator of creation of quantum D0-brane. On
the other hand, it is nothing else but an operator of quantum field of
collective excitations on Fermi-sea i.e. closed string tachyon.

In this paper we suggest a new framework which allows explicit realization of
the setup~\cite{Verl1,Verl2} sketched above. Namely, we study $c=1$ MQM in the
operator Hamiltonian formalism rather than usual functional integral approach.
We construct the $N$-fermion wave functions of eigen-values (free
fermions). Since the system is free $N$ fermions just occupy 
first $N$ one-particle states up to some Fermi-level.
Its shape is given by the one-particle
wave function on the highest level i.e. with
Fermi energy and therefore in quasi-classical limit it
is just a classical trajectory -- a hyperbola.  Next we describe
(following~\cite{AKK}) the system in the non-trivial closed string background,
which can be described as a deformation of the Fermi-sea. For that
we introduce the two-dimensional reformulation of~\cite{AKK}, which
makes the construction of~\cite{AKK} more explicit and well-defined.  We show
that introduction of a classical eigen-value into the system (suggested
in~\cite{Verl1} as addition of the classical D-brane) is equivalent to 
some particular deformation. This is explicit realization of action of
eqs.~(\ref{eq:18})--(\ref{eq:25}) of ~\cite{Verl1,Verl2}.
 
We see that the presence of the brane (that corresponds to moving classical
fermion) induces ripples on the Fermi surface.  In other words, the brane does
change close-string background.  Our approach allows us not only to see this
effect but to describe it quantitatively.  We explicitly present the shape of
the perturbed Fermi-sea even for finite perturbations.  This is a very
detailed picture of the brane back-reaction.  The interaction between the
brane and close string background may exhibit critical behavior.  For example,
the Fermi-sea surface can form cusps even at finite distances between the
surface of the Fermi sea and classical eigen-value. The appearance of the
cusps indicates that the classical description of the system breaks down.

Then we turn to the case of a quantum version of this additional eigen-value.
We present a second quantized bosonic field $\phi$ that creates and destroys
pulses on the Fermi-sea surface thus explicitly realizing the bosonization
formula~(\ref{Ver.bos}).  Vertex operator $e^\phi$ creates $\n$-fermion state
out of $N$ fermion states.  CFT of this field $\phi$ describes quantum
fluctuations of the shape of the Fermi-sea. These are collective bosonic
excitations, which are related to the original fermionic degrees of freedom
only via non-linear bosonization procedure. One-point function of this vertex
operator between vacuum and Dirichlet boundary state is equal to wave-function
of one-fermion state. This further supports the identification of the quantum
eigen-value with D0 brane, suggested in~\cite{Verl2}.

There is one more aspect of our paper that should be stressed. Although all
our results can be obtained from purely MQM point of view they have close
analogies in another system. It is that of 2d electrons in magnetic field,
studied extensively in Quantum Hall effect (QHE). This system is related to
the two-dimensional formulation of the MQM described above via analytic
continuation. Essentially, quantum mechanics of the fermions in the inverse
harmonic potential becomes that of the normal harmonic potential, which, in
particular makes Fermi sea compact. In this work we will often refer to these
cases as \emph{non-compact} and \emph{compact} one correspondingly.  Both
systems has similar phenomena.  Introduction of point-like magnetic fluxes
(counterparts of classical branes) outside the QHE droplet curves its shape.
Bosonic edge excitations, propagating on the surface of Fermi sea (QHE
droplet), which play the important role in the physics of QHE, are collective
excitations of the system of electrons.

This close relation gives the possibility to gain intuition about one system
from the other one. It should be interesting to try to interpret the analog of
phenomena of one system after analytic continuation.

\section{The $c=1$ Matrix Quantum Mechanics}
\label{mqm}

In this section we briefly review results about $c=1$ MQM as well as results
of~\cite{AKK}, which we are going to need later.

The $c=1$ Matrix Quantum mechanics is defined by the following partition
function (we choose here to follow Klebanov \& Gross \cite{Kleb.Gross})
\begin{eqnarray}
  \label{eq:60}
  Z_N(\beta) = \int D^{N^2} M\, e^{-\beta\int(\frac12 {\dot M}^2 +
    U(M))} 
\end{eqnarray}
The path integral is taken over all $N\times N$ Hermitian matrices.  To reduce
the number of degrees of freedom in the integral one can diagonalize the
matrix $\Phi$:
\begin{eqnarray}
\Phi(t) = \Omega^{+}(t) \Lambda(t) \Omega(t)
\end{eqnarray}
by some unitary transformation $\Omega$. Here
$\Lambda=diag(\lambda_1,\dots,\lambda_N)$.
The integration over $\Omega$ can be taken explicitly using
the Itzykson and Zuber formula \cite{Itz.Zub}
\begin{eqnarray}
\int D\Omega\, e^{(\Tr(\Omega A \Omega^{+} B))}
= \frac{{\rm Det}\: e^{(a_i b_i)}}{\Delta(a_i)\Delta(b_i)}
\end{eqnarray}
and one  is left only with the integral over eigen values
of matrix $\Phi$
\begin{equation}
Z_N(\beta) = \int \prod_{i=1}^N d\lambda_i(t)\,
\Delta_N(t_2)\Delta_N(t_1)
\exp\left(-\beta \int\limits_{t_1}^{t_2} dt \sum_i(\frac12{\dot \lambda}_i^2 + U(\lambda_i)\right)
\end{equation}
This partition sum defines a system with Hamiltonian
\begin{eqnarray}
  H = \beta \sum_{i=1}^N
  \left\{-\frac{1}{\beta^2} \frac{\partial^2}{\partial \lambda_i^2} +
    U(\lambda_i) \right\}
\end{eqnarray}
that acts on antisymmetrized wave-functions $\Psi_N(\lambda_1\dots\lambda_N) =
\Delta_N(\lambda) \Phi(\lambda_1\dots\lambda_N)$, where symmetric function
$\Phi(\lambda_1\dots\lambda_N)$ is the singlet sector of the wave-function
$\Psi(M_{ij})$ of matrix model. Thus it describes the system of $N$ decoupled
fermions. In the second quantized formalism it can be written as
\begin{eqnarray}
H = \beta \int d\lambda
\left\{
\frac 1 {\beta^2} \partial_\lambda \zeta^{+}
\partial_\lambda \zeta + U(\lambda) \zeta^{+}\zeta
\right\} \label{sq.H}
\end{eqnarray}
where $\zeta$ is a fermionic field.  Let us consider the potential
$U(\lambda)= 1/4 \lambda^2(2-\lambda)^2$ In the double scaling limit in which
$N$ goes to infinity and $\beta$ goes to some critical value such that the
difference between Fermi-sea level and the top of the potential is kept fixed
one gets\footnote{We introduce yet another fermionic field $\psi$ to stress
  that it is not $\zeta$ of eq.~(\ref{sq.H}). It is related to $\zeta$ by
  rescaling.}
\begin{eqnarray}
H = \int dx \left\{ \frac12 \partial_x \psi^+ \partial_x \psi
- \frac{x^2}{2} \psi^+ \psi
\right\} \label{sq.dsl}
\end{eqnarray}
This is a famous result that the $c=1$ matrix quantum mechanics is
described in the double scaling limit as a system of free fermions
in the upside-down oscillator potential. Going back to first quantized
language the Hamiltonian of eq.~(\ref{sq.dsl}) is
\begin{eqnarray}
\label{eq:3}
H = \frac 12 \sum_{i=1}^N ({\hat p}_i^2 - {\hat x}_i^2 )
\end{eqnarray}
For finite $N$ a final change of variables 
\begin{eqnarray}
{\hat x}_\pm=\frac{{\hat x} \pm {\hat p}}{\sqrt 2}~~~~~~~~~~~[{\hat x}_+,{\hat x}_-]=-i \label{def.xpm}
\end{eqnarray}
brings this upside-down oscillator Hamiltonian to a simple form
\begin{eqnarray}
H_0 = \frac 12({\hat x}_+{\hat x}_-+{\hat x}_-{\hat x}_+) \label{ham.xpm}
\end{eqnarray}
Since ${\hat x}_+,{\hat x}_-$ have canonical commutation relations it is easy
to find eigen-functions and spectrum of this Hamiltonian.  In the $x_-$
representation the operator ${\hat x}_- = x_-$ and ${\hat x}_+ = - i
\partial_{x_-}$. Or equivalently one could switch to $x_+$ representation
${\hat x}_- = i \partial_{x_+}$, ${\hat x}_+ = x_+$.  The solutions to
eigen-function equations are\footnote{\label{fn:1}For simplicity we present
  the results in $x_+$ representation only. $x_-$ representation is totally
  similar. For details see~\cite{AKK}.}
\begin{equation}
  \label{eq:4}
  \psi^E_+(x_+) = \frac1{\sqrt{2\pi}}
e^{-\frac i{2}\phi_0(E)} 
x_+^{i E - \frac 12}
\end{equation}
All scattering data is encoded in the phase factor $\phi_0(E)$ which can be
determined from the simple orthonormality condition
\begin{eqnarray}
\bra{\psi_-^{E_1}}{\hat S}\ket{\psi_+^{E_2}} =  
 \delta(E_1-E_2)
\label{norm1}
\end{eqnarray}
Here $\hat S =\cos( x_+x_-) $ is an operator of Fourier transform between
representations $x_+$ and $x_-$ (for details, see~\cite{AKK}). 

This is one-particle data but since the system is free it is sufficient to
describe the whole MQM.  As it is well known the system is dual to the 2d
string theory in the linearly growing dilaton background.  In the above
duality closed string degrees of freedom correspond to pulses propagating on
the Fermi-sea surface of 1d fermions \cite{Polchinski}.  We would like to
study the property of the system in the presence of closed strings
excitations. Thus we need to change d=1 MQM in such way that it is still
equivalent to system of free fermions but Fermi-sea of these fermions should
be curved.  On string theory side it means to have a time-dependent background
produced by inserting closed-string tachyon vertex operators.  The way to do
it on MQM side was suggested in~\cite{AKK}.  Below we review it briefly.  The
tachyonic excitations are introduced in QM of fermions by assuming that after
this non-trivial background is switched on, the initial wave-functions change
to
\begin{eqnarray}
\label{eq:5}
\Psi_+^E(x_+) =
e^{- i\phi_+(x_+;E)} \psi^E_+(x_{+}) 
\end{eqnarray}
They differ from $\psi^E_+$ in eq.~(\ref{eq:4}) only by a
phase factor $\phi_+$.
This phase can be represented by three terms with different asymptotics.
\begin{eqnarray}
\phi_+(x_+;E) = V_+(x_+) + \frac12\phi(E) + v_+(x_+;E) \label{1dphase}
\end{eqnarray}
The function $V_+(x_+)$ is finite at $x_+=0$ and thus
has only positive powers of $x_+$. The term $v_+(x_+;E)$
is finite at $x_+=\infty$. The shape of $V_+(x_+)$
fixes the perturbation of the background. It is parametrized by
\begin{eqnarray}
V_+(x_+) = R \sum_{k\ge1} t_{k} x^{k/R}_+
\end{eqnarray}
Here R is the radius of compactification.\footnote{In what follows we are
  going to work at self-dual radius $R=1$.}  The rest of the $\phi_+$ (that is
$\phi(E)$ and $v_+$) is uniquely determined by couplings $t_k$ and $E$ from
the orthonormality condition used in eq.(\ref{norm1}):
\begin{eqnarray}
\label{ort1}
\int\int_0^\infty dx_+dx_-{\bar \Psi^{E_1}_-(x_-)}
\cos( x_+x_-)
\Psi_+^{E_2}(x_+) = \delta(E_1 - E_2) 
\end{eqnarray}
In the quasi-classical limit $E_+\rightarrow\infty$
this condition gives
\begin{eqnarray}
x_+x_- = - E_+ + x_+ \partial \phi_+(x_+) \label{fermi.tk}
\end{eqnarray}
Since $x_+x_-$ is just an energy, the above expression
shows that inclusion of $\phi_+$ curves
the shape of the Fermi-sea indeed.
Knowing the wave-functions one can place the system in a box.
and obtain the density distribution of energy in the system. This is
sufficient to calculate free energy of the system. It was done in \cite{AKK}
and the answer was shown to be a tau-function of Toda lattice hierarchy.

The important conceptual problem
comes from an attempt to find an explicit Hamiltonian $H$ that 
possesses the set of eigen-functions eq.~(\ref{eq:5}). Acting naively we
get
\begin{eqnarray}
\label{eq:6}
H \Psi_+ = \Bigl(H_0 + x_+ \partial \phi_+(x_+;E)\Bigr)\Psi_+=E\Psi_+
\end{eqnarray}
As it was just said the dependence on energy of the phase $\phi_+$ is
determined by condition eq.~(\ref{ort1}). Once $t_k$'s are switched on,
$\partial \phi_+$ in~(\ref{eq:6}) is fixed and non-trivial function of E. That
means that Hamiltonian depends on its own eigen value.  In attempt to give
sense to the equation (\ref{eq:6}), it was re-written in \cite{AKK} as
\begin{eqnarray}
H = H_0 + x_+ \partial \phi_+(x_+;H)
\end{eqnarray}
that implicitly defines $H$. 
It does not seem clear at all how to give 
precise mathematical meaning to such equation.
In any case, it is not possible to solve it. 

The difficulty in defining a Hamiltonian obscures the description of
non-trivial tachyonic background.  It seems that without some explanation of
this problem, the ansatz eq.~(\ref{eq:5}), from which all the information was
derived in~\cite{AKK}, remains a physical conjecture, rather than fully
derived solution.  We discuss possible resolution of this issue in the next
chapter.

\section{2D electrons in magnetic field}
\label{sec:magn-2d}

Let us first consider a close relative to Hamiltonian of the MQM, namely, just
a simple one-dimensional harmonic oscillator. The MQM Hamiltonian can be
obtained from it by analytic continuation. We will do it at the end of this
chapter.  The analog of representation of~(\ref{ham.xpm}) in this case is
simply\footnote{It is convenient to keep parameter $B$ for future use.}
\begin{eqnarray}
H = B (a^+ a + \frac12),\qquad [a,a^{+}]=1 \label{ham.osc}
\end{eqnarray}
Instead of representing $a^+=x$, $a=\partial_x$ as we did before we now
suggest another representation.  Let us realize the operators $a,a^+$ as
operators acting on a space of functions on 2d plane rather than on
$\mathbb{R}^1$. Introduce complex coordinates $z,\bz$ on this plane:
\begin{eqnarray}
a = -{\bar \partial} - \frac12 z , \qquad
a^+ = \partial - \frac12 \bz \label{ham.aa+}
\end{eqnarray}
In what follows we are going to show that this trick which may seem as very
formal in the beginning will turn out to have rich physical implication.
After plugging in definition~(\ref{ham.aa+}) into~(\ref{ham.osc}) one can
recognize a Hamiltonian of 2d electrons into constant magnetic field
represented in complex coordinates. Indeed, the usual fermionic Hamiltonian in
constant magnetic field (we set mass and charge of the electron to 1)
\begin{eqnarray}
H = (p - A)^2 \label{Hx},   ~~~ A^i = (\frac{B}{2} y, -\frac{B}{2} x)
\end{eqnarray}
after defining complex variables as
\begin{eqnarray}
z = \sqrt{B/2}(x + i y), \qquad
{\bar z} = \sqrt{B/2}(x - i y) \label{def.zbz}
\end{eqnarray}
assumes the form eq.~(\ref{ham.osc}) with $a,a^+$ defined
in eq.~(\ref{ham.aa+}).
Although the Hamiltonian has a form of 1d harmonic
oscillator the system now is two-dimensional and as a consequence
each energy level is degenerate. This is due to emerged in 2 dimensions
rotational invariance of the system.
Each degenerate state on one level is labeled by eigen-values of
angular  momentum. For example, on the lowest Landau level (LLL)
states are defined by
\begin{eqnarray}
  \label{eq:70}
  a \ket{\psi_0} = ({\bar \partial} + \frac12 z)\ket{\psi_0} = 0 \\
  \psi^\0(z,\bz) = f(z) e^{-\frac12 z\bz} 
  \label{fun.hol}
\end{eqnarray}
The holomorphic function $f(z)$ is arbitrary, showing infinite degeneracy (see
footnote~\vref{fn:3} for discussion of its origin).  To parametrized the space
of solutions of functions on degenerate level one can introduce a basis of
eigen-functions of angular momentum operator~$J$
\begin{eqnarray}
  \label{eq:58}
  J = z \partial - {\bar z} {\bar \partial}
\end{eqnarray}
In this case
\begin{eqnarray}
\psi_n^\0(z,\bz) =\frac{z^n}{\sqrt{\pi n!}} e^{-\frac12 z\bz}~~~~ n=0,1,2\dots \label{2dfun}
\end{eqnarray}
What is less known is that the angular momentum has oscillator representation
too.  Let us define the second set of oscillator operators (called sometimes
\emph{the operators of magnetic translations},
c.f.~\cite{w-inf})\footnote{Problem in the uniform magnetic field should be
  translational invariant. This is why there should be set of two operators,
  commuting with the Hamiltonian~(\ref{Hz}). However, gauge choice breaks this
  invariance, and translations should be accompanied by gauge transformations.
  Therefore operators of magnetic translations $b$ and $b^+$ do not commute
  and are responsible for the infinite degeneracy of LLL.\label{fn:3}}
\begin{eqnarray}
  \label{eq:37}
  b = \partial + \frac12 \bz , \qquad
  b^+ = -{\bar \partial} + \frac12 z 
\end{eqnarray} 
Together with $a,a^+$ they have usual commutator relations of two
independent oscillators
\begin{eqnarray}
[a,a^{+}] = 1, ~~~ [b,b^{+}] = 1,~~~~[a,b] = 0
\end{eqnarray}
The Hamiltonian and angular momentum now take the form
\begin{eqnarray}
H = B( a^{+} a + \frac 12) \label{Hz}\\
J = b^{+} b - a^{+} a
\end{eqnarray}
We see that there are two oscillators in the system.  The energy is defined
with respect to one of them.  The second one specifies an ordering within each
infinitely degenerate Landau level. Namely, one can define ``the real vacuum''
state that is annihilated by both $a$ and $b$ operators. All other states can
be created acting on this state by various powers of $a^{+}$ and $b^+$.  In
this way within each Landau level there is a natural ordering of the states in
consecutive eigen-values of angular momentum.\footnote{Note that in the theory
  Quantum Hall Effect one usually assumes the existence of shallow confining
  potential depending on $J$ which lifts the degeneracy.  Its role is to
  ensure the order of filling one-particle states in many-particle system.
  \label{fn:7}}

If we restrict ourselves to the first Landau level, we see that $J=b^+ b$ and
the problem of finding spectrum of operator of angular momentum is the same as
finding a spectrum of a Hamiltonian of 1d harmonic oscillator.  We just
observed that a system of 2d fermions in constant magnetic field confined to
first energy level is equivalent to 1d oscillator. The wave-functions on the
first Landau level are completely determined by the holomorphic function
$f(z)$ in eq.~(\ref{fun.hol}).  The factor $\exp(-\frac 12 z\bz)$ is common
for all of them and thus all wave-functions can be represented on a space of
holomorphic function, i.e. using $f(z)$ in place of~(\ref{fun.hol}). This can
be done if one also changes the scalar product, namely, instead of
\begin{equation}
  \label{eq:55}
  \sprod{\psi_1} {\psi_2} = \int d^2 z \overline{ \strut\psi_1(z,\bz)}
  \psi_2(z,\bz)
\end{equation}
one writes
\begin{eqnarray}
  \sprod{f_1}{f_2} = \int d\mu(z,\bz) \,\overline{\strut f_1(z)} f_2(z) \label{s.prod}
\end{eqnarray}
with measure $d\mu(z,\bz)$ defined via
\begin{equation}
  \label{eq:56}
  d\mu(z,\bz) \equiv dz d\bz \,e^{-z\bz}
\end{equation}
We will make use of the holomorphic representation in section~\ref{qD}.

\subsection{The $c=1$ Matrix Quantum Mechanics as 2d system}
\label{analyt}
In order to extend the equivalence to the case of MQM with inverse oscillator
potential we have to perform analytical continuation. Our goal is to get such
representation of 2d system that after its reduction to zero-energy level all
wave-functions will be the functions of one real variable, not holomorphic
function as before.\footnote{ The measure in the scalar product
  eq.~(\ref{s.prod}) will be changed accordingly.}

Let us first write the Hamiltonian of inverse oscillator potential eq.~(\ref{ham.xpm})
one more time. We change notations to make the comparison with what was done
above more explicit. The Hamiltonian is
\begin{eqnarray}
&&H = B (p^2 - x^2)/2 = B ( a^+ a - \frac i2) \label{ham.inv}\\
&&a = (p - x)/\sqrt 2, ~~~~ a^{+} = (p + x)/\sqrt 2, ~~~~~ [a,a^+] = - i
\end{eqnarray}
As before, consider the two-dimensional system of electrons in magnetic field
eq.~(\ref{Hx}), do ``Wick''-rotation by changing $y \rightarrow it$ and
consider imaginary magnetic field $\CE=iB$. The vector-potential is changed to
$A=(\frac B2 it, - \frac {B}2 x)$ and new Hamiltonian is
\begin{eqnarray}
  H = (p_x - A_x)^2 +(p_y-A_y)^2&\to& (-i \p_x + \frac B2 it)^2 + ( -i \p_{it}
  + \frac {B}2 x)^2 \label{Ham.iB} \\
  &=&(-i\p_x+ \frac \CE 2 t)^2 -(  -i\p_{t}  -  \frac {\CE}2 x)^2\label{eq:78}
\end{eqnarray}
After such continuation, equation $H\psi=0$ with $H$ given by eq.~(\ref{eq:78})
looks like a massless Klein-Gordon equation in $1+1$ dimensions $(x,t)$ in the
background of constant electric field:
\begin{equation}
  \label{eq:79}
   [-(p_t - A_t)^2 + (p_x - A_x)^2 ]\psi(x,t) = 0
\end{equation}
where $A_\mu \equiv (A_t,A_x) = (-\CE x/2,\CE t/2)$ and thus electric field
$F_{tx} \equiv \p_t A_x - \p_x A_t = \CE$.

Thus, we have shown that after the analytic continuation the
system~(\ref{ham.osc}) of two-dimensional (non-relativistic) particles in the
uniform magnetic field $B$ looks like a relativistic $1+1$ dimensional system
in the uniform electric field $\CE$.

It is well-known that there can be various treatments of this problem. For
example, one may take the solutions of eq.~(\ref{eq:79}) to be the analytic
continuation of solutions~(\ref{fun.hol}),~(\ref{2dfun}). Or one can consider
usual scattering states of the Klein-Gordon eq.~(\ref{eq:79}), with
corresponding scalar product. These two situations describe totally different
physics. For our purpose (namely, reconstruct $c=1$ MQM in singlet sector
starting from analytically continued problem in magnetic field) we will adopt
the first approach. 

Let us introduce new two-dimensional coordinates $z_+,z_-$ which are
analytical continuation of $z,\bz$ in eq.~(\ref{def.zbz}). After
``Wick''-rotation they are
\begin{eqnarray}
z_+ = \sqrt{\CE/2}(x + t), \qquad
z_- = \sqrt{\CE/2}(x - t) \label{def.z+z-}
\end{eqnarray}
In these variables the Hamiltonian eq.~(\ref{Ham.iB}) is
\begin{eqnarray}
  &&H = \CE ( a^+ a - \frac i2), ~~~~ [a,a^+] = - i\label{eq:80}\\
  &&a = -i \partial_+ - \frac 12 z_-, ~~~~ 
  a^+ = -i \partial_- +\frac 12 z_+ 
\end{eqnarray}
where $\partial_\pm = \partial_{z_\pm}$.  This is indeed the Hamiltonian of
the inverse oscillator~(\ref{ham.inv}) but in two-dimensional representation.
Notice, that Klein-Gordon equation~(\ref{eq:79}), from the point of view of
two-dimensional representation~(\ref{def.z+z-}) looks like a condition of the
``lowest Landau level'' $H \ket{\psi_0} = 0$, with $H$ given by
eq.~(\ref{eq:80}).

The wave-functions of the ``ground state'' $a \ket{\psi_0} = 0$ are
\begin{eqnarray}
  \psi_0(z_+,z_-) = f(z_-) e^{\frac {i}2 z_+z_-} 
  \label{fun.zpm}
\end{eqnarray}
In the space of all function $f(z_-)$ we can introduce a basis of
eigen-functions of the (analytically continued) angular momentum\footnote{The
  term $i/2$ in eq.~(\ref{eq:177}) would correspond to $1/2$ of the compact
  case. In that case it is not necessary and can be introduced or not
  depending on physical interpretation of 1d problem.  Here it is important to
  have it in order for $J$ to have eigen-functions~(\ref{eq:178}) with real
  eigen-values. For more detailed discussion of the issue see
  e.g.~\cite{Pioline}.}
\begin{eqnarray}
\label{eq:518}
&&J = b^+ b - a^+ a - \frac i2,\qquad [b,b^+]=-i \label{eq:177} \\
&&b = i \partial_- + \frac 12 z_+ , \qquad 
b^+ = i \partial_+ - \frac 12 z_- 
\end{eqnarray} 
Now operator $J$ acts again on states on the ``lowest Landau level'' in the
same way as 1d Hamiltonian. The LLL states in the basis of eigen-functions of
operator $J$ assume the form
\begin{eqnarray}
\label{0inv}
\psi_E(z_-) = z_-^{i\,E - \frac12} e^{\frac i2 z_+z_-} \label{eq:178}
\end{eqnarray}
We defined the eigen value of $J$ by $E$ to stress that it is not necessarily
an integer as in oscillator case.  The reasoning that forced $E$ to be integer
$n$ in eq.~(\ref{2dfun}) was single-valuedness of the wave-function $z^n
\exp(-\frac12z\bz)$ on a complex plane.  Now $z_-$ is real and $E$ can be any
real number.\footnote{Analytically continued Hamiltonian~(\ref{eq:78}) can
  arise in many different physical situations.  For example, this system is
  known to describe both open strings in the constant electric field and
  twisted sector of the closed strings in Milne
  universe~\cite{Nekrasov,Pioline}. However there is a difference.  While in
  the former case spectrum of operator $J$ is continuous, in the latter it is
  discrete. From this perspective our problem looks like an open string theory
  in the constant electric field.} In MQM system $E$ is an energy that has
continuous spectrum as it should be in unbounded potential.

We can make the relation to MQM system even more explicit by
representing $J=b^+b-i/2$ as acting on space of functions
$f(z_-)$ instead of functions $\psi_0$, $\psi_0=f(z_-) \exp(iz_+z_-/2)$
\begin{eqnarray}
b &\rightarrow& e^{-\frac i2 z_+ z_-}\, b\, e^{\frac i2 z_+ z_-} = i\partial_- \\
b^+ &\rightarrow& e^{-\frac i2 z_+ z_-}\, b^+ \, e^{\frac i2 z_+ z_-} = - z_- \\
J = b^+ b -\frac i2&\rightarrow& - i z_- \partial_- - \frac i2
\end{eqnarray}
This is nothing else but ``$x_-$'' representation of the MQM Hamiltonian
eq.~(\ref{ham.xpm}).  Thus we showed that ``Wick''-rotated 2d system of
electrons in magnetic field on the zero-energy level is equivalent to 1d
MQM.\footnote{What used to be the norm eq.~(\ref{s.prod}) for states on LLL
  for electrons in magnetic field turns after analytic continuation to
\begin{eqnarray}
\sprod{f_1}{f_2} = \int dz_+dz_- e^{iz_+z_-} f_1(z_-) f_2(z_+) \nonumber
\end{eqnarray}
which is equivalent to~(\ref{norm1}), (\ref{ort1}).}

\subsection{Relativistic interpretation}
\label{sec:relative}
We have just reformulated the system of one-dimensional electrons in quadratic
potential as two-dimensional electrons in constant magnetic field, confined to
the first Landau level.  The requirement for 2d fermions to stay on the lowest
energy level can be viewed as a constraint $H=0$.  Therefore, instead of
solving an eigenvalue problem $H\psi =E\psi$ in one dimension, we are solving
equivalent $H\psi=0$ problem in two dimensions. $2d$ system is more general
because the ordering in the space of its solutions can be introduced in many
different ways, e.g.  using any operator, commuting with $H$. By choosing angular
momentum as such operator we define what will play the role of the energy from
$1d$ point of view. Choosing different operators we would arrive to different
$1d$ systems using the procedure described above.

The situation is reminiscent of the problem of choosing particular time in
general relativity.  If one wants to quantize canonically, say, relativistic
particle, it could be done in two ways.  First, make $n+1$ decomposition,
choose the coordinate time $t$ and build $n$-dimensional Hamiltonian.  The
problem to solve will be just $H\psi =E\psi$ where $E$ is actually $p_t$, the
energy defined with respect to time $t$. Otherwise, one could take
relativistic Lagrangian of the form $L=\sqrt{g_{ab}\dot x^a \dot x^b}$, where
dot means derivative with respect to some additional parameter $s$.  Using $s$
as Hamiltonian time, one can build $n+1$ dimensional Hamiltonian system. It is
well known that this system will have Hamiltonian equal to zero and additional
constraint quadratic in momenta. This is due to the fact that relativistic
action is re-parameterization invariant and the Lagrangian is a homogeneous
function of degree one with respect to velocities.  On quantum mechanical
level, one should impose Hamiltonian constraint on the wave functions and
solve the equation $H\psi=0$, where $H$ now is quadratic (in momenta)
constraint.  Any solution of this equation defines some physical state. But to
fully recover the $n$-dimensional eigen-value problem or $n$-dimensional
dynamics, obtained in the first approach, one should add some more
information. Indeed, one should explicitly specify which time is used for the
definition of the energy. To do this, one needs to find some time-like vector
and build corresponding quantum operator. This operator should commute with
the constraint.  Then, the solutions of the constraint will be ordered with
respect to the eigen-values of this operator. In this way one recovers again
$n$-dimensional eigen-values problem in the space of physical states, i.e. in
the space of the solutions of the constraint $H\psi =0$.

The relation between one and two-dimensional systems in our case is formally
exactly of this type.  Namely, imposing the lowest Landau level condition, we
limit the space of functions to holomorphic ones only (see eq.~(\ref{2dfun})).
In this way we effectively reduce the system to one dimension.  This can be
also seen from the fact that one of the two pairs of canonic operators is
frozen under this condition. Then, choosing two-dimensional angular momentum
as an ordering operator, we define one dimensional energy.  This explains the
nature of this relation. It is very interesting to see directly what is the
physical meaning of ``relativistic'' two-dimensional system of electrons from
the point of view of matrix model and string theory.  We will discuss this
issue elsewhere. Now we will just use the formal equivalence between two
formulations of the problem. We will see that it is in two-dimensional
formulation where tachyonic background can be introduced in Hamiltonian more
naturally and that the nature of the problem with its introduction in
two-dimensional system will become clear.\footnote{While this paper was in its
  long preparation, the paper~\cite{Strominger} appeared. In that paper the
  different choice of time in string theory corresponded to time evolution in
  dual MQM generated by different Hamiltonians. This is parallel to what was
  said in this section where we suggested that MQM theory itself (a theory of
  free 1d fermions) was a 2d relativistic theory with the specific choice of
  time already made.  Another choice of time would result in another operator
  of the same MQM generating new time evolution.  The important difference is
  that we fix the time in MQM while in~\cite{Strominger} it is done on the
  string theory side.  It would be interesting to see directly how choices of
  time in both relativistic theories are related.}

\subsection{Introducing non-trivial background. }
\label{tk}

Having built one-particle wave-functions~(\ref{2dfun}) let us turn to the
system of $N$ free fermions on the lowest Landau level. As mentioned in
footnote~\vref{fn:7}, first $N$ consecutive one-particle eigen-states of
angular momentum are occupied and the antisymmetrized function is (we
re-introduced $\hbar$ in this formula)\footnote{Note that notation $\Psi_N$
  will be reserved for the non-normalized function. We will specify
  normalization explicitly, when needed.}
\begin{eqnarray}
  \label{eq:69}
  \Psi(z_i,\bz_i) = \Delta_N(z) \exp\left(-\frac 1{2\hbar} \sum_{i=1}^N z_i
    \bz_i\right) 
\end{eqnarray}
One can show (see e.g.~\cite{cappelli}) that in the so-called dispersionless
limit, when the number of fermions $N\rightarrow \infty$ and $\hbar
\rightarrow 0$, such that $N \hbar =t_0=const$, the density of the fermions
approaches radial step function, which is equal to one inside a circle with
area $t_0$ and zero outside.  The ground state of fermions is described
therefore by a circular droplet of incompressible liquid.  The shape of the
droplet is defined by the trajectory of the fermion with the highest angular
momentum.

Now we turn to the case of MQM. The role of the angular momentum is played by
one-dimensional energy (in a sense of
section~\ref{sec:relative},~\ref{analyt}).  The Fermi-sea of the fermions is a
``droplet'' in the 2d phase space. The shape of the surface of Fermi-sea is
determined by the phase-space trajectory of the fermion with the highest
energy.  Fermions move in inverse quadratic potential and thus their
trajectories are hyperbolae rather than circles and the droplet has the form of
non-compact Fermi-sea, as we expect from the MQM.

To describe dual to MQM closed strings degrees of freedom we have to allow for
the pulses propagating on Fermi surface as suggested in \cite{Polchinski}.
Thus we need to consider the system with deformed Fermi sea, like it was done
in~\cite{AKK} and reviewed in the section~\ref{mqm}.

To do this we notice the following.  First Landau level in magnetic system is
infinitely degenerate and one can choose there any orthonormal basis.  As
discussed in section~\ref{sec:relative}, we can choose any operator, commuting
with the Hamiltonian~(\ref{Hz}) to define the ordering of the basis states.
After analytic continuation such operator will serve as one-dimensional
energy. It means, for example, that when we build ground state of the system
of $N$ particles, we will define it as a Slater determinant of the first
consecutive $N$ eigen-functions of this operator. We want to describe the
ground state corresponding to perturbed Fermi-sea (as in section~\ref{mqm}).
This essentially means that we should choose the basis of functions, similar
to those of~\cite{AKK} (given by eq.~(\ref{eq:5})).  We will do it in two
steps.  First, we notice that if we multiply any function of the lowest Landau
Level by the fixed factor $e^{\omega(z)}$ with $\omega(z)$ being the
\emph{entire} function, given by
\begin{eqnarray}
  \label{eq:7}
  \omega(z)= \sum_{k\ge 1} t_k z^k, ~~~~ V(z,\bz) = 2 \Re\omega \label{tk.hall}
\end{eqnarray}
it is still the function of LLL.\footnote{
  Notice, that it is important
  for the function $\omega(z)$ to be an entire function, i.e. for the
  series~(\ref{tk.hall}) to have infinite radius of convergence. If the radius
  of convergence were finite, this would imply that new function is \emph{not}
  a solution of equation~(\ref{eq:70}) anymore. If one changes wave function
  multiplying it by $\exp{g(z, \bar z)}$, it would satisfy an equation with
  \emph{different} gauge potential $A_\bz$, such that magnetic field $B$
  (physical observable) has changed by $\delta B=2\p \bar\p (g(z,\bar z)+\bar
  g(z, \bar z))$.
  
  In the case of~(\ref{tk.hall}) $g(z,\bz)=\omega(z)$ and naively $\delta B$
  is equal to zero. However, if one has a finite radius of convergence
  in~(\ref{tk.hall}) then it can change $B(z, \bz)$.  For example, function $
  \omega(z) = \sum_a q_a \log (z-\zeta_a)$ can be represented
  as~(\ref{tk.hall}) only for $|z|<\min_a(|\zeta_a|)$ with $\t k = \sum_{a}
  \frac{q_a}{k \zeta_a^k}$. For such $\omega(z)$ functions~(\ref{2dfuntk})
  satisfies the equation~(\ref{eq:70}) with \emph{different} gauge potential
  $A_\bz=z+\bar\p\omega(z)$, such that magnetic field $B$ has changed by $
  \delta B(z,\bz) = \sum_{a=1}^M q_a \delta^\2 (z-\zeta_a)$ which means that
  the magnetic field has been changed by a number of point-like fluxes
  inserted far from the droplet.  This situation is physically different from
  the one considered above and it was first analyzed in~\cite{qhe}. For
  example, in case of $\omega(z)$ given by (sum of) logarithm, building of
  orthogonal polynomials $p_n(z)$~(\ref{2dfuntk})--(\ref{eq:8}) may work
  differently. Namely, in case of integer (negative) fluxes one-particle
  states~(\ref{2dfuntk}) do not change, i.e. $p_n(z)e^{\omega(z)}=z^n$,
  however first $Q$ states $n=0,\dots,Q-1$ (where $Q = \sum_a |q_a|$) are
  absent.  Thus, such $\omega(z)$ corresponds to creating a hole of size $Q\h$
  in the middle of the (round) droplet of size $N\h$ (provided, of course,
  that $Q\ll N$).}
Each basis state that was an eigen state with some angular momentum is
transformed to a new one. These new states can be numbered by eigen-values of
some new operator. Let us try to build the corresponding operator.  To do
this, we build a new canonic pair of operators:
\begin{eqnarray}
  b = \partial + \frac12 \bz - \sum_{k=1}^\infty k t_k z^{k-1},\quad
  b^+ = -{\bar \partial} + \frac12 z,\quad [b,b^+]=1 \label{newb}
\end{eqnarray}
The basis of one-particle wave-functions will be defined by
\begin{eqnarray}
  &&b \psi_{0,0} = (\partial + \frac12 \bz - \sum_{i=1}^k k t_k z^{k-1}) 
  \psi_{0,0} = 0,\quad \psi_{0,0}=e^{-\frac12 z\bz + \omega(z)}\label{eq:72}\\
&&\psi_{0,n} = (b^+)^n  \psi_{0,0}
= z^n e^{-\frac12 z\bz + \omega(z)} \label{2d0}
\end{eqnarray}
where still
\begin{equation}
  \label{eq:41}
  a \psi_{0,n} = (- {\bar \partial} - \frac12 z ) \psi_{0,n} = 0
\end{equation}
Notice, that compared to~\cite{AKK} we have introduced here only a part of
potential containing positive degrees of $z$.  Next we are going to show how
a part with negative degrees appears.

Having introduced new basis~(\ref{2d0}), one should check that these
wave-functions are orthogonal.  Indeed, they are not.  At first glance this is
surprising, because the functions $\psi_{0,n}$ are eigen functions of operator
$\CJ =b^+b$. The reason is that operator $\CJ$ is not hermitian anymore for
$b^+,b$ defined in eq.~(\ref{newb}) (if we had introduced $b^+$ hermitian
conjugated to the new $b$, the corresponding functions~(\ref{2d0}) would not
have belonged to the first Landau level).  Therefore, operator $\CJ$ is not a
good choice to define a one-dimensional energy for ``deformed case''. We need
to find a linear transformation of our basis functions~(\ref{2d0}) to make
them orthogonal. The result will have the form\footnote{We restore $\h$ in the
  rest of this section, as it it is important for taking dispersionless
  limit.}$^{,}$\footnote{Operator of one-dimensional energy is defined by its
  eigen-functions~(\ref{2dfuntk}).  Later, when we build $p_n(z)$ explicitly,
  we will see that~(\ref{2dfuntk}) are nothing else but Backer-Akhiezer
  functions from the theory of integrable systems. It is known there that the
  corresponding operator is a complicated non-local object.}
\begin{eqnarray}
\psi_n = p_n(z) e^{-\frac1\h\left(\frac{z\bz}{2} + \omega(z)\right)} \label{2dfuntk}
\end{eqnarray}
where $p_n$ are polynomials of degree $n$: $p_n(z)=z^n + \dots$.
The orthogonality condition that defines
$p_n(z)$ is
\begin{eqnarray}
\label{ort}
\int d^2z~ \psi_N(z)\bar\psi_M(\bz) =
\int d^2z~ e^{-\frac{z\bz}\h} z^N \bz^M e^{\,\frac1\h\sumk (t_k z^k +\bar t_k
  \bz^k)}e^{\frac1\h(v(z,N)+\bar v(\bz,M))} =h_M\delta_{MN}  
\end{eqnarray}
Here we introduced $v(z,N)$ such that 
\begin{eqnarray}
\label{eq:8}
 v(z,N)\equiv \h \log(p_N(z)/z^N)\equiv-\sum_{k\ge 1} \frac{v_k(N)}{ k z^k} 
\end{eqnarray}
It is easy to see that $v(z,N)$ depends on $t_k$ and on the value of angular
momentum $N$ which is to become an energy.  For a reference, in the case of
zero-potential $p_N(z) = z^N$ and $v=0$.  We see that the orthogonality
condition~(\ref{ort}) is absolutely analogous to the condition~(\ref{ort1})
which appeared in~\cite{AKK} and was the reason to have restrictions on $v(z)$
making it dependent in $t_k$ and $E$.  If we tried to obtain the wave
function~(\ref{2dfuntk}) from the one-dimensional problem, we would encounter
the same problem: one-dimensional Hamiltonian $H=a a^{+}+\frac{1}2$ would be
shifted by $\partial \left(\omega(z)+v(z,E)\right)$ after introduction of
$t_k$ into wave-functions and become energy-dependent.  As we have seen, from
the two-dimensional point of view the negative part of the potential is
obtained in a different way (by orthogonalization procedure) and it does not
cause any problem.

A physical explanation why the two-dimensional formalism fits better for the
description of tachyonic background could come from its relativistic
interpretation suggested in section~\ref{sec:relative}.  
In this interpretation it is
natural that once one introduces non-trivial time-dependent background, it
becomes hard to use coordinate time for Hamiltonian formalism. Relativistic
Hamiltonian formalism with background independent Wheeler-DeWitt type equation
is still applicable.

In the large $N$ limit the ground state defines a droplet of incompressible liquid
as was said before and the orthogonality condition defines the shape of the
droplet. Namely,  for $N=M$ in the eq.~(\ref{ort}) we can use saddle point
approximation to see that for $N\h = \t0$ the droplet has a boundary, given by
the equation 
\begin{equation}
\bz= S(z)=\sum_{k\ge1} k t_k
z^{k-1}+\frac{t_0}z+\sum_{k\ge1} v_k z^{-k-1}\equiv S_+(z)+S_-(z)\label{eq:9}
\end{equation}
Under certain conditions on the function $S(z)$ (for details
see~\cite{davis,wz,kkmwz} or appendix~\ref{sec:schwarz}), it defines analytic
curve in the complex plane. Function $S(z)$ is called \emph{Schwarz function}
of the curve. Positive part of its Laurent series~(\ref{eq:9}) $S_+(z)$
contains $t_k$, which are the same as in eq.~(\ref{ort}). Coefficients $v_k$
of the negative part $S_-(z)$ are the dispersionless limit of $v_k(N)$ in
eq.~(\ref{eq:8}) and are the functions of $t_k$ and $t_0=N\hbar$. In the
inverse potential case the analog of this curve is the shape of Fermi sea
specified by equation $x_+x_-=\mu +x_+\p_+ \phi(x_+)$. We see that by the same
reason as it was in~\cite{AKK}, namely from the same orthogonality
condition~(\ref{ort}), we can derive the shape of the Fermi sea. In the
compact case it is such that harmonic moments of the droplet\footnote{See
  appendix~\ref{sec:schwarz} or~\cite{wz,kkmwz} for definitions of harmonic
  moments.} are equal to the coupling constants $t_k$.

Let us consider the simplest example, illustrating the above procedure. We
choose $\omega(z)$ to be a linear function $\omega(z)=\t1 z$. Then, one can
check that the set of orthogonal polynomials is
\begin{equation}
  \label{eq:73}
  p_n(z) = (z-\tb1)^n
\end{equation}
where $\tb1$ is a complex conjugate of $\t1$. Function $v(z,N)$ is defined in
this case as
\begin{equation}
  \label{eq:74}
  v(z,N) = \h N\log\left(1-\frac{\tb1}z\right) = -\sumk\frac{\h N\tb1^k}{k z^k}
\end{equation}
Taking the dispersionless limit $N\to\infty$, $\h\to0$, $N\h = \t0$, we obtain
\begin{equation}
  \label{eq:75}
  v_k(N) \to {\t0\tb1^k}
\end{equation}
This should be compared with the Schwarz function with non-zero $\t1$, given
by
\begin{equation}
  \label{eq:76}
  S(z) = \t1 + \frac{\t0}{z-\tb1} = \t1 + \frac{\t0}z + \sumk\frac{\t0\tb1^k}{z^{k+1}}
\end{equation}
We see, that indeed, limit of $v_k(N)$ coincides with $v_k$ of Schwarz
function.

\section{D-branes and closed strings} 

The formalism developed above suggests new realization of objects, discussed
in~\cite{Verl1,Verl2} and briefly reviewed in~(\ref{part.N})--(\ref{Ver.bos}).
Namely, we will give a meaning to the classical and quantum eigen-values
of~\cite{Verl1,Verl2} and their interaction with the Fermi-sea. We will
construct a local fermion operator, creating eigen-value at the point $\zeta$,
and show how it can be described in terms of fluctuations of the shape of the
Fermi-sea (which are interpreted as closed strings excitations). Then we will
try to define precise meaning of classical eigen-value and in what sense one
can think of it as of D-brane.

\subsection{Quantum eigen-value}
\label{qD}

We begin with the quantum treatment of the problem of interaction of Fermi-sea
with the single eigen-value.  To do this, let us make the following trick
first.  The main difference between the perturbed and unperturbed wave
functions is that basis functions on the LLL (monomials $z^n$) become
orthogonal polynomials $p_n(z)$.  Let us try to build these polynomials. To do
this let us consider first the $N$-particle function in the perturbed case (we
re-introduce $\hbar$ in this section and switch to the holomorphic
representation, which amount to removing all $e^{-\frac1{2\h}|z|^2}$ and using
measure~(\ref{eq:56}) instead of $d^2 z$ in scalar products):
\begin{eqnarray}
\Psi_N(z_i) = \frac1{\sqrt{N!}}\Delta_N(z_i) 
\prod\limits_{i=1}^N e^{ \frac{\omega(z_i)}\h}
\label{eq:14}
\end{eqnarray}
The antisymmetrized product of $p_n(z)$ gave rise to the Van der Mode
determinant as in the constant magnetic field case. Note that if we started
with non-orthogonal wave functions, we would get the same answer.\footnote{The
  only thing which is important for that is that $p_n(z) = z^n +
  \dots$\label{fn:2} } The norm of the state~(\ref{eq:14}) as a function of
$t_k$ is given by
\begin{equation}
  \label{eq:15}
  \tau_N(\t k,\tb k) 
     =\frac{1}{N!}
  \int |\Delta_N(z)|^{2}\prod_{i=1}^N e^{\frac {V(z_i,\bar z_i)}\h}d\mu(z_i ,\zb i)
\end{equation}
It is easy to see that\footnote{
  Note that $\Psi_\n$ in (\ref{eq:39}),(\ref{N+1}) are \emph{not} normalized!}
\begin{eqnarray}
  \label{eq:39}
  \Psi_\n(z_1,\dots,z_N,\zeta|\t k ) = e^{
    \frac{\omega(\zeta)}\h  }\Delta_N(z)\prod_{i=1}^N (z_i-\zeta)  
  e^{\frac1\hbar\sum\limits_{i=1}^N \omega(z_i)}
\end{eqnarray}
Note, that following~\cite{Verl1,Verl2} factor $\prod(\z i -\zeta)$ in
eq.~(\ref{eq:39}) can be interpreted as interaction of $N$ quantum
eigen-values with $N+1$-st classical, in position $\zeta$. Then l.h.s. of
eq.~(\ref{eq:39}) can be interpreted as an $N$-particle wave-function in the
different background, given by new $\t k$, shifted by $\h /(k\zeta^k)$.
Notice, that one may think about $\h$ as string coupling constant $g_s=\h$.
Then the new background is described by the $\tilde{\t k} = \t
k-\frac{g_s}{k\zeta^k}$.  Thus, in the language of~\cite{Verl1,Verl2}
introduction of one more eigen-value $\zeta$ is equivalent to the change of
background (described in the shift of $\t k$). We will return to the
discussion of this issue in section~\ref{cD}.

We emphasized explicitly in the argument of the wave-function in~(\ref{eq:39})
that it depends on $\t k$. The reason for that is that even if we are
interested in the wave functions in trivial background $\Psi_N(z|0)\equiv
\Psi_N(z)$, to define the action bosonic operators on them we have first to
deform Fermi sea by introducing non-zero $\t k$'s, act on them with the
operators, and only then put all $\t k$'s equal to zero. In this sense $\t
k$'s play the role of classical sources of quantum field theory. As a result
we get
\begin{eqnarray}
  \Psi_\n(z_1,\dots,z_N,\zeta|0)=\left[
    e^{\frac{\hat \phi(\zeta)}{\h}} \Psi_N(z_i|\t k)\right]\biggr|_{\t k = 0}
  \label{N+1}
\end{eqnarray}
where we have introduced the operators:
\begin{equation}
  D(z)=\sum_{k\geq 1}\frac{z^{-k}}{k}\frac{\p}{\p t_k},\ \ \ \ \ 
  \hat\phi(\zeta)= \h N\,\log \zeta + \omega(\zeta) - \h^2 D(\zeta) 
\label{eq:20}
\end{equation}

Next, we are going to build a single-particle wave-function~(\ref{2dfuntk})
for $n=N$.  Using many-particle functions we can represent it as
$\psi_N(\zeta) = \la N,\zeta \ket{N+1}$, where $\ket{N,\zeta}\equiv \ket
N\otimes \ket \zeta$.  It means that
\begin{equation}
  \label{eq:16}
  \psi_N(\zeta) = \int \prod_{k=1}^Nd\mu( \z k,\zb k)
  \frac{\bigbar\Psi_N(\bz_1,\dots,\bz_N) \Psi_{N+1}(\z1,\dots,\z
    N,\zeta)}{\sqrt{\tau_N \tau_{\n}}} 
\end{equation} 
or explicitly\footnote{Note that the functions $\psi_N$ built in this way are
  automatically orthogonal.  It is easy to see from the fact that if we
  started from orthogonal functions we would get them again at the end and
  that many particle wave-function has the same form regardless of what basis
  we started with (see footnote~\vref{fn:2}).  Therefore this procedure is an
  orthogonalization procedure.  One can easily extract from~(\ref{eq:19}) the
  expression for the polynomials $p_n(z)$:
  $$
  p_n(z) = z^n \frac{e^{- \h D(z)}\tau_n(\t k,\tb k)}{
    \sqrt{\tau_n\tau_{n+1}}} = z^n \frac{\tau_n(\t k-\frac{\h }{kz^{k}},\tb
    k)} {\sqrt{\tau_n(t,\bar t) \tau_{n+1}(t,\bar t) }}
  $$
  From this expression it is not obvious to see that $p_n(z)$ indeed
  defines a polynomial. This is a non-trivial result known from the theory of
  integrable system.}
\begin{equation}
  \label{eq:17}
  \psi_N(\zeta) = \frac
  {e^{\frac{\omega(\zeta)}\hbar}}
  {\sqrt{\tau_N \tau_{\n}}}
  \int \prod_{k=1}^N d\mu(\z k,\zb 
  k)  |\prod_{i<j}^N (\z i - \z j)|^2  (\zeta-\z k)  e^{
    \frac{ V(\z k. \zb k)}\h}
\end{equation}
As in eq.~(\ref{N+1}), we can rewrite expression~(\ref{eq:17}) trying to take
all $\zeta$ dependence out of integral (c.f.~\cite{qhe}):
\begin{equation}
  \label{eq:19}
  \psi_N(\zeta ) = \frac{e^{\frac 1\h(\h N \log \zeta + {\omega(\zeta)} - 
      \h^2 D(\zeta))}\tau_N(\t k,\tb k)}{\sqrt{\tau_N \tau_{\n}}}
  \equiv 
  \frac
  {e^{\frac{\hat\phi(\zeta)}{\h}}\tau_N(\t k,\tb k)}{\sqrt{\tau_N \tau_{\n}}}
\end{equation}
To further interpret expression~(\ref{eq:19}) and to make explicit its
connection to D-branes (according to conjecture of~\cite{Verl1,Verl2}), one
needs to take large $N$ limit in it. Namely, we take $N\to\infty$ and $\h\to
0$, while keeping their product $\t0 = N \h$ finite.  This gives (for
simplicity, we are computing $\psi(\zeta)$ on the ``round'' background: $\t k
= \tb k=0$)\footnote{Term $\p_{\t0}$ is responsible for the normalization and
  can be included in the definition of operator $\hat\phi$.}
\begin{equation}
  \label{eq:21}
  \psi_N(\zeta)\to\psi(\zeta) \sim \left.
    \frac{e^{ \frac{t_0}\h\log \zeta + \frac1\h\omega(\zeta) -
        \h D(\zeta)}\tau(\t0,\t k, \tb k)}{\tau(\t0,\t k,\tb k)}
  \right|_{\t k = \tb k = 0}
\end{equation}
with
\begin{equation}
  \label{eq:22}
  \tau_N(\t k,\tb k|\h) \to e^{\h^{-2} F(\t0,\t k,\tb k)}
\end{equation}
and $F=\log\tau$. In the future $\tau$ will always denote this particular
tau-function -- the dispersionless limit of 2D Toda hierarchy.

The same formula~(\ref{eq:21}), relating fermionic wave function to the
bosonic field via $\tau$ function, exists also for the non-compact case of
original $c=1$ string theory (inverse potential) (see e.g.~\cite{kostov}, eq.
4.20). This is clearly a bosonization formula.  Notice that, unlike
in~\cite{KMS}, this bosonization is not asymptotic, it is valid everywhere in
the large $N$ limit.

Several words should be said to identify the mathematical meaning of functions
$F$ in~(\ref{eq:22}) as well as $\tau_N$.  Notice first that $\tau_N$
in~(\ref{eq:15}) is the tau-function of (dispersionful) 2D Toda hierarchy
(see, e.g.~\cite{moerbeke}).  There are many ways to see it. For example,
eq.~(\ref{eq:15}) can be identified with the partition sum of the normal
matrix model, if one identifies $\z i$'s with the eigenvalues~\cite{zaboron}.
This partition sum is known to give the particular tau-function of 2D Toda,
obeying additional constraints, called \emph{string equation}. Then the
eq.~(\ref{eq:22}) defines function $F$ -- the logarithm of the tau-function of
the\emph{ dispersionless} 2D Toda hierarchy~\cite{takasaki-takebe}.

\subsection{CFT of edge excitations and boundary state}
\label{sec:cft2}

In this section we are going to show, how one can interpret formulae of the
preceding section in terms of conformal field theory.

First of all, notice that representation~(\ref{eq:20}) shows that $\hat\phi$
can be thought of as an operator for the bosonic field in two dimensions.
Indeed, let us denote
\begin{eqnarray}
  \label{ak}
  &&a_k= \hbar \partial_{t_k},\qquad  a_{k}^{+}= \frac k\hbar t_k,
  \qquad  [a_k,a^+_n] = k \delta_{k,n}
  \\ 
  &&\hat \phi(z)= \h \left[N\log(z) +  \sum_{k>0}
    \left(\frac {z^k}k a_k^+ - \frac {z^{-k}}k a_k \right)\right]
  \label{eq:57}
\end{eqnarray}
Thus $\hat \phi$ is an just an operator of free scalar field.  Defined above
$a_k$'s act on a space of functions of infinite number of variables $f(\t
k)\equiv f(\t1\dots\t k\dots)$.  To make it a Hilbert space, one needs to
define proper scalar product on that space.  Let $\ket{g}$ denote the state
that corresponds to a function of infinite variables $g(t_k)$ and $\bra f$ is
the Hermitian conjugation of the state, defined by $f(\t k)$.  We define
scalar product as
\begin{equation}
  \label{eq:59}
  \sprod{f}{g} \equiv \left(\bar f(\p_{t_k}) g(t_k)\strut\right)|_{t_k = 0}
\end{equation}
One can easily see that such scalar product is (1) linear and (2) positively
defined.\footnote{This is obviously true for the polynomial functions. We
  understand all functions of our Hilbert space as given by their Taylor
  series and thus being completion of the space of polynomial functions.} %
Thus the bra-vector $\bra{f}$, corresponding to the state $\ket f$, which is
described by the function $f(t_k)$ is represented by the linear functional
$\bar f(\p_{t_k})$ acting on the functions of $t_k$.  Both left and right
vacua $\ket0$ and $\bra0$ correspond to $f(t)=1$.

Let us try to see what is the physical meaning of such states $f(\t k)$ in terms
of original quantum mechanics. It is well-known that in the dispersionless
limit electron droplet of compact case (or Fermi-sea of non-compact case) is
incompressible. As it was discussed in section~\ref{tk}, states, corresponding
to incompressible deformations, can be described via wave
functions~(\ref{2dfuntk}) (or~(\ref{eq:14}) on the level of $N$-particle
states), where arbitrary $\t k$'s parameterize arbitrary shape of Fermi-sea.
Notice, that because we are interested in incompressible deformations only, it
is enough for us to consider functions of the form~(\ref{eq:14}). Every
element of our state of incompressible deformations is parameterized by set of
$\t k$'s.

As it is discussed in details in~\cite{qhe-cft}, to make this considerations
explicit, one can switch to the so-called \emph{\trep}.  Namely, take any
state $\ket\Phi$ in the $N$-particle sector, represented by the
(antisymmetric) function of $N$ variables $\Phi_N(\z1\dots\z N)$. In the
future we will often call this function \emph{$z$-representation} of state
$\ket\Phi$ to distinguish it from \emph{\trep}, which is defined in the the
following way
\begin{equation}
  \label{eq:1061}
  \Phi(\s k) \equiv \sprod{\s1\dots\s k\dots}{\Phi} =
  \int\prod_{i=1}^N d\mu(\z i,\zb i)\,\overline{\strut \Psi}_N(\zb1\dots\zb N|\s k)
  \Phi_N(\z1\dots \z N)
\end{equation}
Function $\overline{\strut \Psi}_N(\zb1\dots\zb N|\s k)$ is a complex
conjugated version of eq.~(\ref{eq:14}), with variables $\tb k$ being
substituted with independent variables $\s k$.  Scalar product of any two
states $\Phi_1(\s k)$ and $\Phi_2(\s k)$ is defined by requirement that it is
the same in \trep, as it is in \zrep. This is the same
definition~(\ref{eq:59}):
\begin{equation}
  \label{eq:65}
  \sprod{\Phi_2}{\Phi_1}_s \equiv \int\prod_{i=1}^N d\mu(\z i,\zb
  i)\,\overline{\strut \Phi}_2(\zb1\dots\zb N) \Phi_1(\z1\dots \z N) =
  \bigbar\Phi_2(\p_{\s k}) \Phi_1(\s k)\Bigr|_{\s k =0}
\end{equation}
Then tau-function $\tau_N(\t k,\s k)$ describes the \trep\ of the
state~(\ref{eq:14}), which we will often call $\ket{\t k}_N$:
\begin{equation}
  \label{eq:1058}
  \tau_N(\t k,\s k)\equiv \sprod{\s1\dots\s k\dots}{\t k}_N =
  \int\prod_{i=1}^N d\mu(\z i,\zb i)\,\overline{\mathstrut \Psi}_N(\zb1\dots\zb N|\s k)
  \Psi_N(\z1\dots \z N|\t k)
\end{equation}
We should stress, that eq.~(\ref{eq:1058}) is very \emph{asymmetric} with
respect to $\t k$ and $\s k$. Namely, one should think of it as a function
of $\s k$, parameterized by the set of (in general complex) numbers $\t k$. 

With these definitions, we can re-write (complex conjugate of)
eq.~(\ref{eq:21}) as matrix element of CFT. Indeed,
\begin{equation}
\label{eq:13}
  \bar\psi_N(\zeta) = 
  \left.\frac{e^{\frac{\hat\phi_s(\zeta)}\h}\tau_N(\t k,\s
      k)}{\sqrt{\strut\tau_N(\t k,\tb k)\tau_\n(\t k,\tb k)}}
  \right|_{\s k = \tb k}  = \frac{\bra{\tb k}
    e^{\frac{\hat\phi_s(\zeta)}\h}\ket{\t k}_s}{\sqrt{\strut\sprod{\tb k}{\tb
        k}_s\sprod{\t k}{\t k}_s}} 
\end{equation}
where state $\bra{\tb k}$ is given by the linear functional:
\begin{equation}
  \label{eq:62}
  \bra{\tb k} = \exp\left(\sum_{k=1}^\infty \tb k \pfrac{\;}{\s
      k}\right) 
\end{equation}
Again, if we are interested in results for the unperturbed Fermi sea and use
$\t k$'s only as sources, we should put $\t k=\tb k=0$
in~(\ref{eq:13}).\footnote{
  In general case of non-zero $\t k$'s it was shown
  in~\cite{qhe} that $|\psi(z)|^2$ is localized on the contour, specified by
  harmonic moments $\t k$ and area $\t0$ (for the procedure of constructing
  the contours via this data see~\cite{wz,kkmwz}):
\begin{displaymath}
  |\psi(z)|^2 \sim \delta(\CC)|w'(z)|
\end{displaymath}
where $w(z)$ is the conformal map from the exterior of the curve to the
exterior of the unit circle.}$^,$\footnote{Normalization in eq.~(\ref{eq:13})
comes from the fact, that we have defined states $\ket{\t k}$
via~(\ref{eq:1058}) using \emph{non-normalized} wave-functions $\Psi_N(z|\t
k)$ and $\bigbar\Psi_N(\bar z| \s k)$.}

Equation~(\ref{eq:13}) is already interpreted as (one point) correlator of
CFT. From it we see, that expectation value of vertex operator
$\exp(\frac{\hat\phi_s(\zeta)}\h)$ can be interpreted as fermion. Namely, can
prove that $e^\frac{\hat\phi(\zeta)}\h$ is \trep\ of the following \emph{local
  fermion operator} $\hat\varPsi^+(\zeta)$:
\begin{equation}
  \label{eq:64}
  \hat\varPsi^+(\zeta) \equiv \sum_{k=0}^\infty \bar\psi_k^\0(\zeta) \hat\CO_k^+
\end{equation}
where $\bar\psi_k^\0$ are complex conjugates of one-particle basis
states~(\ref{2dfun}), and operators $\hat\CO_k^+$ create particles in the
state~(\ref{2dfun}). (Of course, this operator may be equivalently defined
with respect to any one-particle basis, and we choose that of~(\ref{2dfun})
only for definitiveness). Then, one can show (again, see~\cite{qhe-cft} for
details) that
\begin{equation}
  \label{eq:61}
  \bra{N+1,\tb k} \hat\varPsi^+(\zeta)\ket {N,\t k} = \bar\psi_N(\zeta)
\end{equation}
where $\psi_N$ is given by~(\ref{eq:21}) or~(\ref{eq:13}). 
Therefore, we have checked on the level of the one-particle expectation value,
that these operators are the same.\footnote{It turns out (see~\cite{qhe-cft}),
  that eq.~(\protect\ref{eq:61}) is enough to prove the equivalence.
  Nevertheless, it is very instructive to calculate explicitly a two-point
  function in both representations, and we are going to do it in the next
  section.}

\subsection{Density correlator and boundary state}
\label{sec:boundary-state}

Finally, we are going to show why it is possible to speak about object,
corresponding to the expectation value of operator~(\ref{eq:64}), as of
D-brane, thus giving confirmation to Verlinde conjecture. Namely, we are going
to show that object $\tau_N$, may be given an interpretation different from
those of~(\ref{eq:1058}). 
To wit, in the large $N$ limit this object may be interpreted as a Dirichlet
boundary state $\bket{D}$.

To show it, let us start from the microscopic density matrix is defined via
\begin{equation}
  \label{eq:66}
    \rho_N(\zeta,\zeta') = \frac1{\tau_{\n}(\t k,\tb k)} \int
    \prod_{k=1}^Nd\mu( \z k,\zb k) 
  {\bigbar\Psi_{N+1}(\bz_1,\dots,\bz_N,\zeta'|\tb k) \Psi_{N+1}(\z1,\dots,\z
    N,\zeta|\t k)}
\end{equation}
(in computations of this section we are going to think of $\t k$ and $\tb k$,
as of complex conjugated variables).  This can be re-written as
\begin{equation}
  \label{eq:67}
  \rho_N(\zeta,\zeta') = \frac{\bra{\t k, N}\hat\varPsi^+(\zeta')\hat\varPsi(\zeta)\ket
  {\t k, N} }{\sprod{\t k,N}{\t k,N}}
\end{equation}
From the previous sections is is obvious that~(\ref{eq:67}) can be computed in
\trep:
\begin{equation}
  \label{eq:68}
   \rho_N(\zeta,\zeta') = \bra{\t
     k,N}e^{\frac{\overleftarrow\phi(\zeta')}\h}e^{\frac{\overrightarrow
       \phi(\zeta)}\h}\ket{\t k, N}_s  
\end{equation}
If one writes down eq.~(\ref{eq:68}), using definitions~(\ref{eq:59}) of scalar
product in \trep, this will be very non-trivial expression, difficult to
compute. However, from explicit \zrep~(\ref{eq:66}) one can see that
\begin{equation}
  \label{eq:71}
  \rho(\zeta,\zeta')=\left.\frac{e^{\frac{\phi(\zeta)+{\bar\phi}(\zeta')}{\h}}\tau_N(\s k,\sb 
    k)}{\tau_\n(\t k,\tb k)}\right|_{\s k =\t k, \sb k = \tb k}
\end{equation}
where operators $\phi(\zeta)$ and $\bar\phi(\zeta')$ are two copies
of~(\ref{eq:57}), acting in variables $\s k$ and $\sb k$ correspondingly.
Thus, we see that one can reinterpret ``chiral operator'' $\phi$, acting on
the left and on the right in eq.~(\ref{eq:68}), as a ``non-chiral'' operator
$\phi+\bar\phi$, acting on the right in eq.~(\ref{eq:71}).

To make this fact more understandable from CFT point of view, let us recall
the following property of the tau-function in eq.~(\ref{eq:71}).  As it was
shown in~\cite{br} one can interpret the dispersionless tau-function of 2D
Toda lattice hierarchy as Dirichlet boundary
state on the arbitrary analytic curve \CC. It is known 
that Dirichlet boundary state on the unit circle is given by
\begin{equation}
  \label{eq:34}
  \bket D = \exp\left(\sum_{k=1}^\infty \frac{a^+_k \bar a^+_k}k\right) \ket 0
  \otimes \ket{\bar 0}
\end{equation}
The analog of the formula on the arbitrary curve \CC, which is related to the
second derivatives of the dispersionless Toda tau-function in view of
works~\cite{wz,kkmwz}, is given by the following equation
\begin{equation}
  \label{eq:36}
  \bket{\CC}=
\exp\left(\frac{a_0^2}{2}\frac{\p^2 F}{\p\t 0^2} +\sum\limits_{k,n=1}^\infty
  \frac{\p^2 F}{\p\t k \p\t n} \,\frac{a^+_k a^+_n}{2 k\,n} + 
  \frac{\p^2 F}{\p\t k \p\bt n} \frac{a^+_k \ba n^+}{2 k\,n} +\frac{\p^2 F}{\p\t k \p\t
    0}\,\frac{a_0 a_k^+}{k}+\,{c.c.}\right)
\ket 0\otimes \ket{\bar 0}
\end{equation}
(complex conjugation in~(\ref{eq:36}) means interchange of $\t k$ with $\bt k$
and $a^+_k$ with $\ba k^+$ correspondingly).  Essentially, every term $a^+_k
a^+_n$ in~(\ref{eq:36}) is contracted with the corresponding second derivative
of dispersionless tau-function. Equation~(\ref{eq:34}) is the particular case
of this construction if one takes into account that for the unit circle one
has
\begin{equation}
  \label{eq:38}
  \frac{\p^2 F}{\p \t k \p\bt k} = k
\end{equation}
with all other second derivatives equal to zero. 

Thus we have shown that
\begin{equation}
  \label{eq:40}
  \rho(\zeta,\zeta')=\bra{\t k}\otimes\bra{\tb k} e^{\phi(\zeta) + \bar
    \phi(\zeta')}\bket{D} 
\end{equation}
Coherent states $\bra{\tb k}$ (and its analog $\bra{\t k}$) are defined
in~(\ref{eq:62}). One may be surprised that the result of eq.~(\ref{eq:71})
(where the whole function $\tau_N(\t k,\tb k)$ seems to play the role), may be
reproduced using~(\ref{eq:36}), which consists of only second derivatives. The
explanation lies in the fact that in the quasi-classical limit
eq.~(\ref{eq:13}) contains only first and second derivatives of tau-function.

We see that wave-function $\psi(z)$ of quantum eigen-value is given by the
matrix element of the bosonic vertex operator $e^\phi$ between the vacuum and
Dirichlet boundary state. This supports the conjecture of~\cite{Verl1,Verl2}
that quantum eigen-value corresponds to $D0$ brane. Although our formalism
sheds some new light on this correspondence, its string theoretical
interpretation requires further clarification.

\subsection{Classical eigen-value}
\label{cD}
Let us now return to the question of an eigen-value located outside Fermi sea,
which was discussed at the beginning following~\cite{Verl1}.  The main idea of
this consideration was that if we add a classical particle located outside the
droplet (Fermi sea), it interacts with the particles in the sea because of
Pauli principle. Therefore, the $N$-particle wave function gets multiplied by
the factor $\prod ( z_i-\zeta (t) )$, where $\zeta (t)$ is the trajectory of
the classical particle.  This means that each one-particle wave function will
change in the presence of classical particle just by adding
\begin{eqnarray}
  \omega(z) = \h\log (z -\zeta(t)) \label{chi.log}
\end{eqnarray}
to potential in the exponent
\begin{eqnarray}
  \label{eq:81}
  \psi_E=z^E e^{-\frac1{2\h}z \bz} \rightarrow z^E  e^{-\frac1{2\h}z \bz +
    \frac1\h\omega(z)}=z^E  e^{-\frac1{2\h} z \bz + \frac1\h\sum t_k z^k} 
\end{eqnarray}
where $t_k=-\frac {\h}{k \zeta^k}$.  From two dimensional point of view the
eq.~(\ref{chi.log}) describes magnetic flux placed at $\zeta(t)$, whose
strength is $\h$.  This is the same modification of background, which we have
already discussed in section~\ref{qD} after eq.~(\ref{eq:39}).  Let us see to
which background it corresponds.  Having wave functions modified like
in~(\ref{eq:81}), we need, as explained in section~\ref{tk}, to make them
orthogonal.  One could think, that this would give rise to the orthogonality
condition~(\ref{ort}) from which we can see in the semi-classical
(dispersionless) limit that the shape of Fermi sea is not a hyperbola (or
circle in compact case), but some non-trivial curve defined by $t_k$. This is
however \emph{not} necessarily the case.  Indeed, as mentioned above, $\t k$'s
obtained from~(\ref{chi.log}) are proportional to $\h$. Thus in the
quasi-classical limit $\h\to0$, they do not change the shape of the Fermi-sea.
On the other hand, we may try to take different limit: take $\h=g_s$ to be
fixed and send $N\to\infty$. In the compact case this would mean that the
droplet is actually growing (with the area proportional to $N$) and we do not
seem to gain much with the deformation with finite size.  However in the
non-compact case we are only interested in the surface of the (infinite)
Fermi-sea and this change is finite and observable.  Thus, the question how
classical eigen-value is defined should be treated more carefully, especially
in the case of compact droplet, where we are interested in the droplets of the
finite size.

To describe interaction of $N$ quantum particles with one classical, one
should build a Slater determinant, consisting of $N$ particles in the ground
state (forming Fermi-sea) and one particle in the state, admitting
quasi-classical description (i.e. sharply localized around its classical
trajectory).  For example, one could put $N+1$-st particle in the state far
above the Fermi-sea level. It is easy to see, however, that the situation in
that case will not be very different from~(\ref{chi.log}). The explanation
lies in the fact that we are working in the dispersionless limit, where number
of particles $N\gg1$.  Then adding just one more particle cannot
macroscopically change the system.  In view of discussion of previous
sections~\ref{qD}--\ref{sec:boundary-state}, we see, that it is more correct
to think about this system as of quantum field theory system, than of quantum
mechanical one. In quantum field theory classical objects are usually consist
of the large number of elementary quantum excitations. Thus it seems that if
we want to have a deformation of the system, which does not disappear in our
limit (where Fermi-sea is defined), we should think about classical
eigen-value as of consisting of \emph{many} fermions. For example, we can put
large number $q\gg1$ in~(\ref{chi.log}). Then we would get the following
deformation of $N$-particle state:
\begin{equation}
  \label{eq:77}
  \Psi_N(\z1,\dots,\z N) \to \Delta_N(z) \prod_{i=1}^N(\z i - \zeta)^q =
  \Delta_N(z) \prod_{i=1}^N e^{\frac1\h\sumk \t k z^k_i}
\end{equation}
where now $\t k = -\frac{\h q}{k \zeta^k}$. To realize flux with the large $q$
in terms of eigen-values, we need, of course, to take a droplet of $q$ of
them. To make this droplet look like a point-like flux, we need to make its
size (given by $q\h$) much less than $N\h$ -- the size of the droplet. Because
in the dispersionless limit we choose $N\h$ (and correspondingly $q\h$) to be
arbitrary numbers, there is no contradiction in fact that $q\h\ll N\h$, but
still finite.

To interpret this classical eigen-value as a D-brane
(following~\cite{Verl1,Verl2}), one should recall the result of
section~\ref{sec:boundary-state}, where we showed that one can associate
Dirichlet boundary state with the (compact) droplet. Thus, indeed, classical
eigen-value in a sense described above (as a droplet of many eigen-values) can
be associated with the D-brane. We are not discussing here the question of
decay of this D-brane.

In this section we have mostly discussed the interpretation of classical
eigen-value in case of compact droplet. The interpretation in non-compact case
may be different, due to the fact that we are only interested in deformations
of vicinity of Fermi-surface. Nevertheless, classical eigen-value is described
as small compact droplet also in this case and results of
section~\ref{sec:boundary-state} are also applicable. Thus we suggest that the
classical D-brane in the non-compact case is a small compact droplet, placed
outside the Fermi-sea and moving along the classical trajectory.  Compare this
with~\cite{sen-mm}, where similar scenario was discussed.

In sections~\ref{sec:1miwa},~\ref{sec:shape-non-compact} we will suggest a
possible scenario, in which small droplet separates off the Fermi-sea (see,
e.g. fig.~\vref{fig:3}) and thus interpret this process as possible creation
of D-brane.

We see that the system of fermions with one classical fermion added is exactly
equivalent to the system of fermions with some collection of pulses
propagating along the Fermi sea.  In other words, D-brane defined as a
classical eigen-value (in open string language so to say) is equivalent to
some particular closed string background.

\subsection{Shape of the droplet in case of one flux}
\label{sec:1miwa}

Now we are going to describe explicitly the shape of Fermi-sea induced by the
classical eigen-value placed outside. In principle, classical
eigen-value should move along the classical trajectory as
in~\cite{Verl1,Verl2}. We will first consider the case when the eigen-value
does not move at all (which is also a classical trajectory, at least in the
absence of confining potential). This problem was solved in~\cite{miwa1}.

As mentioned already (c.f. end of the section~\ref{tk} or
appendix~\ref{sec:schwarz}), to describe an analytic curve on the complex
plane one may use the Schwarz function $S(z)$. In the current case of
shape defined by one classical eigen-value, we know the positive part of the
Schwarz function. It is given by
\begin{equation}
  \label{eq:10}
  S_+(z) = \sum_{k\ge1}k \t k z^{k-1} = \frac q{z-\zeta(t)}
\end{equation}
(flux strength $q$ is introduced here for future convenience). To describe the
shape of the Fermi sea one needs to reconstruct negative part of Schwarz
function $S_-(z)$. In principle, as we have seen, it is determined from
$S_+(z)$ by orthogonality condition~(\ref{ort}). However, in case of
quasi-classical limit it is much easier to use geometric construction
of~\cite{wz,kkmwz}. This will help us to guess the correct $S_-(z)$ without
going through the actual computations.\footnote{In general, to find
  coefficients of $S_-(z)$ from $t_k$ we would have to build (dispersionless)
  tau-function of two dimensional Toda lattice hierarchy $F(t)$ and calculate
  $v_k =\p_{t_k} F$.}

There are many ways to describe an analytic curve in the plane. One of them is
to specify the conformal transformation from an exterior of, say, unit circle
to that of the analytic curve. The existence of such map $z(w)$ is guaranteed
by Riemann mapping theorem. The natural question is -- what is the relation
between the two approaches -- Schwarz function and conformal mapping from the
unit circle. The answer is the following (for details
see~\cite{davis,wz,kkmwz}). Given conformal map $z(w)$ for $|w|\ge1$ we can
construct function $\bz(w^{-1})$ in the same domain via
\begin{equation}
  \label{eq:11}
  \bz(w^{-1}) = \overline{\strut z(w)}\Bigr|_{\bw = w^{-1}}
\end{equation}
Then Schwarz function is given by
\begin{equation}
  \label{eq:12}
  S(z(w)) = \bz(w^{-1})
\end{equation}
Putting it differently, parameterization $z(w)$ and $\bz(w^{-1})$ solve
identically unitarity condition~(\ref{eq:32}).

Our strategy is to guess a conformal map such that equation~(\ref{eq:12}) will
reproduce the correct $S_+(z)$~(\ref{eq:10}). Then eqs.~(\ref{eq:11})
and~(\ref{eq:12}) will automatically give the correct Schwarz function.  The
resulting  conformal map is:
\begin{equation}
  \label{eq:24}
  z(w)=w+\frac{A w}{a(w-a)},\quad|a|<1
\end{equation}
\begin{figure}[htb]
  \centering{
    \epsffile{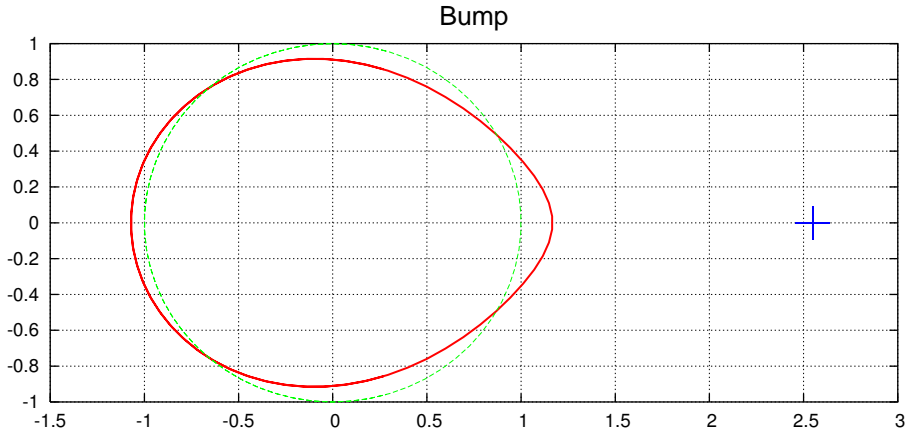}}
  \caption{\small Curve~(\protect\ref{eq:24}), perturbed from initial unit
    circle (thin dotted line, in green, where available) by the flux, marked
    by (blue) cross on the right of the curve.}
  \label{fig:1}
\end{figure}
The computation of $\t k$, corresponding to this curve is done in
appendix~\ref{sec:miwa1}.  One sees that indeed $\t k = \frac{q}{k\zeta^k}$,
where position of the flux $\zeta$ and its strength $q$ are given in terms of
$a$ and $A$ by
\begin{equation}
  \label{eq:35}
\zeta = \frac{1}{\bar a}+\frac{A}{a(1-|a|^2)},\quad 
q = \frac{{\bar  A}}{{\bar a}^2} - \frac {|A|^2}{(1-|a|^2)^2}
\end{equation}
\begin{figure}[htb]
  \centering{
    \epsffile{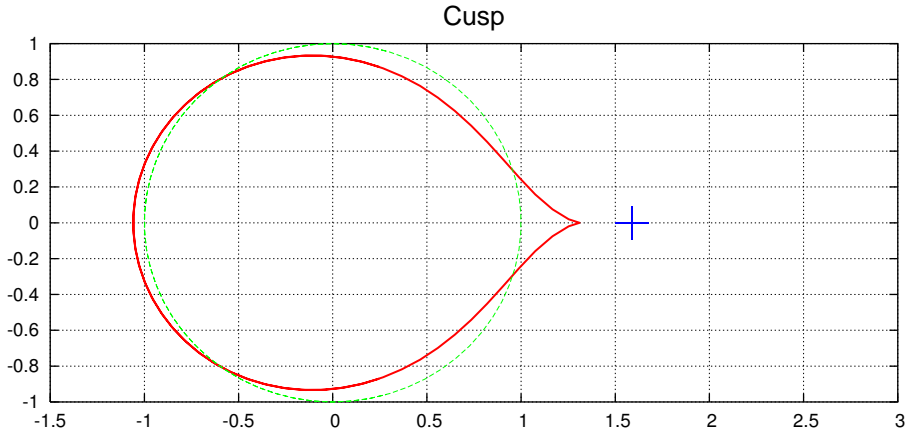}}
  \caption{\small As flux (blue cross) moves closer to the curve (as compared to 
    the figure~\protect\vref{fig:1}) curve develops a cusp (corresponding to
    $z'(w)=0$ in eq.~(\protect\ref{eq:24})). }
\label{fig:2}
\end{figure}
One may easily see that for $A=0$ curve~(\ref{eq:24}) becomes a unit
circle.\footnote{One may also send $a\to0$. Then we get $\zeta\to\infty$,
  however, to be able to take this limit, one should also send $A/a\to 0$.
  Then we also restore the unit circle.}
The shape of the curve starts to change as one increases $a$ (keeping $A$
fixed for simplicity). It develops a bump in the direction of the flux (that
is, the bump is centered along the ray, connecting origin with the position of
the flux). It is shown on the figure~\vref{fig:1}. If we would move flux
adiabatically around the curve, the bump would follow. This is the explicit
realization of the equivalence between the (quasi)classical eigen-value (with
position $\zeta$) and deformation of the Fermi-sea of the
eigen-values~\cite{Verl1,Verl2}.

It is interesting to notice that the change of the shape of the curve
depending on the trajectory of the classical eigen-value exhibits critical
behavior. Namely, as one moves it closer and closer to the curve (i.e.
decreases $\zeta$), bump not only grows but eventually becomes singular -- the
curve develops a cusp while the classical eigen-value is still the finite
distance away from it (see figure~\vref{fig:2}).

\begin{figure}[htb]
  \centering
    \includegraphics{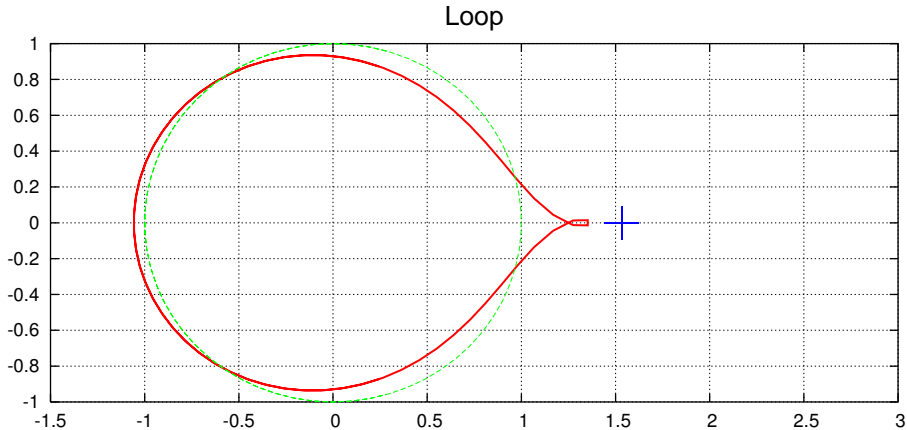}
  \caption{\small If one continues movement of the flux (blue cross) toward
    the curve (compared to fig.~\protect\vref{fig:1}, and
    fig.~\protect\vref{fig:2}, which are all done at the same scale) the curve
    self-intersects. Probably this should be considered as a creation of small
    separate droplet, but this is impossible to see in the quasi-classical
    approximation we are using in this paper.}
\label{fig:3}
\end{figure}

This means that quasi-classical approximation is not applicable anymore. In
this critical region quasi-classical approximation (dispersionless limit) is
not valid anymore, and sub-leading effects become important. At this point the
system changes drastically. The easiest way to see it, is to continue to move
flux closer to the droplet. Then the curve would self-intersect, developing
essentially the region with negative area (see figure~\ref{fig:3}).
Of course, it is not obvious that one can believe the classical analysis in
the self-intersecting region.  These features can also be seen analytically
from eq.~(\ref{eq:24}) and we refer the reader to the
appendix~\ref{sec:miwa1}.

\subsection{Shape of the non-compact Fermi sea in the presence of classical eigen-value}
\label{sec:shape-non-compact}

Similar to the previous section phenomena take place in the non-compact case.
Indeed, the shape of Fermi sea is described by
\begin{equation}
  \label{eq:43}
  x_+ x_- = \mu + x_+S_+(x_+) = \mu + \sum_{k=1}^\infty k \t k x_+^k +
  \CO(\frac 1{x_+})
\end{equation}
Choosing as in the compact case $t_k$'s to be of the form $\frac q{k
  \zeta^k_+}$ we come to the situation similar to~(\ref{eq:10}):
\begin{equation}
  \label{eq:44}
  x_+x_- = \mu + \frac {q x_+}{x_+ - \zeta_+} +   \CO(\frac 1{x_+})
\end{equation}
and similar representation for the $x_-$ representation with (in general
different) coordinate of the flux $\zeta_-$. As in section~\ref{sec:1miwa}, we
can use a trick to find a form of the Fermi surface without finding explicitly
$v_{\pm k}$. Namely, we find a parameterization $x_+(\omega)$, $x_-(\omega)$
which would satisfy eqs.~(\ref{eq:43})--(\ref{eq:44}) identically. In
non-compact case it was used in~\cite{kostov}. There it was possible to find
an ansatz for $x_\pm(\omega)$ because there was only finite number of $t_{\pm
  k}$. In the present case the number of $t_k$'s is infinite, but their
particular form makes the procedure to be possible again.

We take the
ansatz to be of the form:
\begin{equation}
  \label{eq:45}
  x_+(\omega) = \omega - \frac{B\omega}{\omega+a_+}; \quad x_-(\omega) =
  \frac1\omega - \frac {B}{1+a_-\omega}
\end{equation}
where both $B$ and $a_\pm$ are positive and $0\le\omega<\infty$. 
Substituting~(\ref{eq:45}) into the eq.~(\ref{eq:43}) one gets:
\begin{equation}
  \label{eq:46}
  \zeta_+ = -\frac 1{a_-} - \frac {B}{1- a_- a_+}; \quad \mu - \frac q{\zeta_+} = 1
  - \frac B{a_+}
\end{equation}
\begin{figure}[htb]
  \centering
  \includegraphics{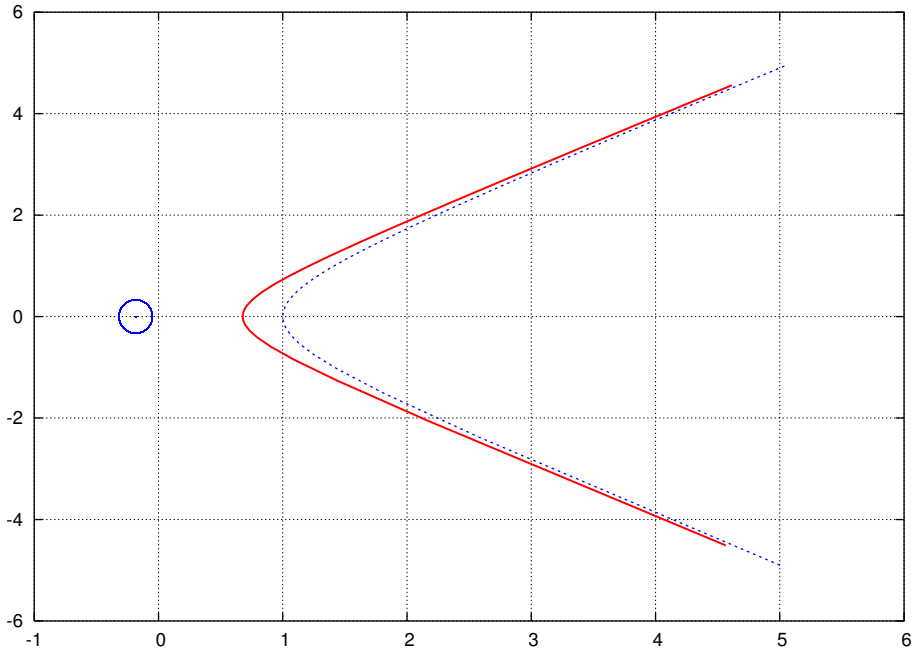}
  \caption{\small Deformation of the (non-compact) Fermi sea in the presence
    of the classical eigen-value. Circle (on the left) marks the position of
    the classical eigen-value. The deformed sea is the solid line. Hyperbola,
    to which it tends at infinities, is thin dotted (blue) line.}
  \label{fig:4}
\end{figure}
There is a particular difference between this case and the compact case of the
previous section~\ref{sec:1miwa}. For example, in non-compact case the whole
Fermi sea was changing with introduction of the flux (see fig.~\vref{fig:1}).
In this case, however, only the area close to the flux gets deformed (see
fig.~\vref{fig:4}), while asymptotically ansatz~(\ref{eq:45}) still gives
\begin{equation}
  \label{eq:47}
  x_+ x_- = \mu'
\end{equation}
-- equation of unperturbed hyperbola. 

Another difference with the compact case is the appearance of the cusp in case
of only $t_{\pm1}$ being non-zero~\cite{kostov}. In the compact case
introduction of $t_1,\bt 1\ne0$ would correspond to the shifted circle with no
critical behavior possible.

Appearance of the cusp is a very intriguing phenomenon and its interpretation
in the context of~\cite{Verl1,Verl2} needs further clarification.

\section{Discussion}
\label{sec:disc}

In this section we briefly repeat the main results of this paper and discuss
its open questions and relation to other works. 

It has already been known for some time (c.f. e.g.~\cite{KKA2}) that there is
an intrinsic relation between $c=1$ MQM and normal matrix model (the latter
describing, in view of~\cite{qhe}, (integer) Quantum Hall system). However,
precise relation between these two systems was not well understood. On the
other hand, although known since early 90s, the fact that $1+1$ dimensional
$c=1$ string theory is described by $0+1$ dimensional $c=1$ matrix
\emph{quantum mechanics} has not received a fully satisfactory explanation.
From yet another point of view, current works on conjecture of~\cite{Verl1}
(see e.g.~\cite{KMS,Verl2} and many other recent references) were based on the
widely accepted fact of duality between MQM and Liouville description of $c=1$
string theory~\cite{Polchinski:94}.

This work adds new ingredients to all of these issues. First, it shows that
results of~\cite{AKK,KKA2} can be better understood if one thinks about MQM as
of analytic continuation of two-dimensional problem of the electron in the
magnetic field. The one-dimensional problem of electrons in inverted harmonic
potential appears from it as restriction to the ``lowest Landau level'', as
discussed in details in section~\ref{analyt}.  Analytically continued problem
is two-dimensional as well, but lives in $1+1$ dimensions, rather than in
$2+0$ as the magnetic system on the lowest Landau level does.  This
construction allows to give microscopic interpretation of the result
of~\cite{KKA2}, where analytic continuation from normal matrix model to MQM
was obtained on the level of partition functions.  In \cite{AKK} the
wave-functions of fermions in non-trivial background, the shape of the
Fermi-sea, and most of the results were obtained through orthogonalization
procedure rather than via solution of an explicit Schroedinger equation. From
the point of view of two-dimensional ($1+1$-dimensional) electrons in the
magnetic (correspondingly, electric) field it corresponds to a choice of
orthogonal basis on the ``lowest Landau level'' defined by equation ${\hat H}
\Psi = 0$. This is in turn equivalent to imposing an additional condition on
the states on the lowest Landau level, namely the condition for them to be the
eigen-states of some operator $L$ commuting with 2d Hamiltonian $\hat H$.
That provides $1+1$-dimensional (rather than $0+1$ dimensional) interpretation
of the fermionic system. The operator $L$ serves as $0+1$ dimensional
Hamiltonian.  Notice, that different choice of operator $L$ (i.e.  different
ordering on the space of solutions) would gives us \emph{different} $0+1$
problem. This resembles a lot the problem of choice of time in general
relativity or any generally covariant theory.  This makes the problem of
string theoretic interpretation of $1+1$ dimensional rather than MQM system
very intriguing.\footnote{Recently, the paper~\cite{Strominger} appeared. In
  that paper the very similar phenomenon is discussed on string theory side --
  different choices of time in $1+1$ dimensional string theory give rise to
  different versions of $0+1$ MQM.  It
  seems very interesting to compare both approaches.} %
Recall, that $1+1$-dimensional system of section~\ref{analyt} has known
interpretation as strings in electric field (see e.g.~\cite{Pioline}).  We
discussed it shortly in the section~\ref{analyt} and plan to return to this
interpretation elsewhere.  This direct string theoretic interpretation of MQM
is in line with recent interpretation of MQM as of world-volume theory of
tachyonic field of D-branes in c=1 string theory~\cite{Verl1}.

Next, we showed that CFT mechanism, developed recently for the Quantum Hall
systems (as described in~\cite{qhe-cft} and in
sections~\ref{sec:cft2}--\ref{sec:boundary-state}) can be applied (in view of
the aforementioned analytic continuation) to the $c=1$ MQM (or its
two-dimensional equivalent) to reproduce all the ingredients of the
conjecture~\cite{Verl1}. In our formalism we are able to identify
``classical'' D-branes with the insertion of the ``classical'' flux and
quantum D-brane with the action of operator $e^{\phi(\zeta)}$, such that
$\psi(\zeta)=\la e^{\phi(\zeta)}\ra$. This shows, that, indeed, D-brane
(described by fermion $\psi(\zeta)$) can be thought of as a soliton of MQM,
built from various closed string excitations -- pulses on the surface of
Fermi-sea, whose creation/annihilation is described by the modes of the
operator of the scalar field $\phi(\zeta)$. On the other hand, classical
D-brane (flux, inserted in the position $\zeta(t)$) can be ``dissolved'' into
the closed string background. We consider also explicit example of such
``correspondence'' between classical D-brane and closed string background and
showed that one can expect critical phenomena there as well as processes of
``creation'' of D-branes.

The question of classical eigen-value needs some further investigation.  We
have mostly discussed the interpretation of classical eigen-value in case of
compact droplet. The interpretation in non-compact case may be different, due
to the fact, that we are only interested in deformations of vicinity of
Fermi-surface. Nevertheless, classical eigen-value is described as small
compact droplet also in this case and results of
section~\ref{sec:boundary-state} are also applicable. Thus we suggest that the
classical D-brane in the non-compact case is a small compact droplet, placed
outside the Fermi-sea and moving along the classical trajectory (For similar
picture see e.g.~\cite{sen-mm}).

We should stress once again, that this is done in the way absolutely different
from the usual approach of comparing results of computations in MQM with those
of Liouville boundary CFT. It may well be that what is described here is a
new, dual description of Liouville CFT (much simpler in some
sense).\footnote{After this paper was finished, reference~\cite{Rastelli}
  appeared, which discussed similar issue of duality between Liouville CFT and
  (Kontsevich) matrix model. For a connection between open string field theory
  and normal matrix model (or, equivalently, Toda Lattice hierarchy) see also
  work~\cite{bkr}.} We leave this investigation for the future.

It is also interesting to note that the description of D-branes, suggested
in this paper is of more general nature. Indeed, recently the
paper~\cite{top-stings} has appeared, which suggests very similar 
realization of classical and quantum D-branes in the context of
topological B-model (where $c=1$ string theory appears as a particular case).

\section*{Acknowledgements}

We would like to thank N.~Nekrasov for fruitful discussions.  A.B. and B.K.
would like to acknowledge warm hospitality of NBI, Copenhagen where this work
was started. A.B. and O.R. also thank AEI, Golm where it was finished. A.B.
acknowledges financial support of Swiss Science Foundation. The work of BK is
supported by German-Israeli-Foundation, GIF grant I-645-130.14/1999

\newpage

\appendix

\section{Schwarz function and harmonic moments of analytic curves}
\label{sec:schwarz}

In this appendix we summarize briefly the results about the description of
curves in terms of harmonic moments. The results can be found in
\cite{minwz,wz,kkmwz} and in book \cite{davis}.

Let \CC\ be an analytic curve in the plane, without self-intersections (for
simplicity), and for simplicity ``sufficiently close to circle'' (see details
in book \cite{davis}). Let denote by $D_+$ interior of curve and by $D_-$ -
its exterior. For simplicity we assume that $z=0\in D_+$ and $z=\infty\in
D_-$. Then one can define a set of harmonic moments of this curve, They are

-- \emph{area}
\begin{equation}
\label{eq:26}
  \t0 = \frac1\pi \int_{D_+}d^2 z =  \frac1{2\pi i}\oint_\CC dz \bz
\end{equation}

-- \emph{moments of exterior}
\begin{equation}
\label{eq:27}
  \t k = \frac1{2\pi i k} \oint_\CC z^{-k} \bz dz
\end{equation}

-- \emph{moments of interior}
\begin{equation}
  \label{eq:28}
   v_k = \frac1{2\pi i } \oint_\CC z^{k} \bz dz
\end{equation}

and also \emph{logarithmic moment}
\begin{equation}
\label{eq:29}
   v_0 = \frac1{\pi } \int_{D_+} \log|z|^2 d^2z
\end{equation}

Then, one can construct the formal Laurent series, called the \emph{Schwarz
  function}
\begin{equation}
  \label{eq:30}
  S(z) = \sumk k \t k z^{k-1} + \frac{t_0}z + \sumk\frac{v_k}{z^{k+1}}
\end{equation}
which defines the curve \CC\ via relation
\begin{equation}
  \label{eq:31}
  \bz = S(z)
\end{equation}
One can show, that eq.~(\ref{eq:30}) has non-zero region of convergence in
strip-like domain around the curve \CC.

This statement can be reversed. Namely, if one has the function $S(z)$,
formally defined via eq.~(\ref{eq:30}) and if this function obeys so called
\emph{unitarity condition} in some region
\begin{equation}
  \label{eq:32}
  \bar S(S(z)) = z
\end{equation}
for all $z$ in the domain of definition, then eq.~(\ref{eq:31}) defines some
curve. Without the condition~(\ref{eq:32}) eq.~(\ref{eq:31}) would define only
a discrete set of points in the complex plane. 

It should be noted, that set $\t k$, $v_k$ is over-determined. One can show that
independent set of variable is all $\t k$ (all $v_k$'s are good set as well) and
that
\begin{equation}
\label{eq:33}
  v_k = \pfrac{F(\t0,\t k,\tb k)}{\t k},\quad k=0,1,\dots
\end{equation}
where $F(t)$ is the logarithm of dispersionless tau-function of 2D Toda
Lattice hierarchy. It is the same tau-function, which appears in the large $N$
limit of normal matrix model.

\section{Construction of the curve with one flux}
\label{sec:miwa1}

We need a Schwarz function, which has one pole in the position $\zeta$ outside
the contour.\footnote{We assume that $S_-(z)$ has no poles.} We are going to
show that one can describe corresponding to such Schwarz function curve by the
following conformal map:
\begin{equation}
  \label{eq:1024}
  z(w)=w+\frac Aa +\frac{A }{w-a} = w + \frac {A}{a(1-\frac aw)}
\end{equation}
One should remember, that this is a map of \emph{exterior} of the unit circle
to the \emph{exterior} of the curve and that it \emph{does not} map interior
of the circle into anything in the conformal way.  To be the conformal map of
exterior of the unit circle, one requires $|z'(w)|\neq 0,\infty$, for $|w|\ge
1$. This leads to the following restrictions on parameters:
\begin{equation}
  \label{eq:48}
  |a| < 1, \quad |a \pm \sqrt{A} | < 1
\end{equation}
Indeed, according to the discussion above (see~(\ref{eq:12})), the Schwarz
function, defined by
\begin{equation}
  \label{eq:1025}
  S(z(w)) =\frac{1}{w}+\frac{\bar A}{\bar a({1}-\bar a w)}
\end{equation}
has a single pole at the point $w_f=\frac{1}{\bar a}$, which means that
position of the flux in the $z$-plane is given by
\begin{equation}
  \label{eq:1035}
  \zeta\equiv z(w_f) = \frac{1}{\bar a}+\frac{A}{a(1-|a|^2)}
\end{equation}
Let us compute $\t k$ for the curve given by conformal
map~(\ref{eq:1024}). According to definition~(\ref{eq:27})
\begin{eqnarray}
t_k = \frac1{2\pi ik} \oint_\CC z^{-k} \bz dz
\end{eqnarray}
We assume that $\oint_\CC \frac {dz}z = - 2\pi i$.
\begin{eqnarray}
  t_k& = &\frac1{2\pi ik} \oint\limits_{|w|=1} z^{-k}(w) \bz(1/w) dz(w) \nonumber\\
  &=&\frac1{2\pi ik} \oint\limits_{|w|=1} \left(w + \frac {A}{a(w-a)}\right)^{-k}
                      \left(\frac1w + \frac {\bar A}{\bar a(1-\bar aw)}\right)
                      \left(1 - \frac A{(w-a)^2}\right) dw  \label{eq:50}\\
                      &=&\frac1{2\pi ik} \oint\limits_{|w|=1} \frac{(w-a)^k}{(w(w-a)
                        + \frac 
                        Aa)^{k}} \left(\frac1w + \frac{\bar A}{\bar a^2} \frac
                        w{\frac1{\bar a}-w}\right) \left(1 - \frac
                        A{(w-a)^2}\right) dw  \nonumber
\end{eqnarray} 
We choose to pick up the poles outside the circle.  There are two poles: at
$w=1/\bar a$ (since $|a|^2 < 1 $) and at $w=\infty$. We believe, that because
of condition~(\ref{eq:48}) poles of quadratic polynomial $w(w-a)+\frac Aa$ are
\emph{inside} the unit circle and thus do not contribute. Also, one can see
that in the case~(\ref{eq:1024}) residue at $w=\infty$ is equal to zero. Thus
we have
\begin{equation}
  \label{eq:49}
  t_k = \frac1k 
  \left(\frac1{\bar a} + \frac A{\frac1{\bar a}-a} + \frac Aa \right)^{-k}
  \frac{{\bar A}}{{\bar a}^2}\left(1 - \frac A{(\frac1{\bar a}-a)^2}\right) =
  \frac 1k \frac q{\zeta^{k}}
\end{equation}
where $\zeta$ is given by eq.~(\ref{eq:1035}) and flux strength $q$ is defined
as
\begin{equation}
  \label{eq:51}
  q = z'({\bar a}^{-1})\res_{w={\bar a}^{-1}} \bz(w)=\frac{{\bar
      A}}{{\bar a}^2}\left(1 - \frac {A\bar a^2}{(1-|a|^2)^2}\right) = \frac{{\bar
      A}}{{\bar a}^2} - \frac {|A|^2}{(1-|a|^2)^2}
\end{equation}
The pole of $S(z)$ and its residue are position and the value of a flux.  If
we want flux to be real, this means that $\bar A/\bar a^2$ should be real as
well.

The area of the droplet can be computed from~(\ref{eq:1024}),~(\ref{eq:1025})
as $t_0=\frac 1{2\pi i}\oint \bar z(w^{-1})dz(w)$ i.e.
\begin{equation}
  t_0= 1 - \frac {|A|^2}{(1-|a|^2)^2}
  \label{eq:52}
\end{equation}
We see, that $\t0$ is always real, regardless the values of parameters $a, A$,
provided that conditions~(\ref{eq:48}) are satisfied.

Now, let us look in more details at the situation when the cusp develops
This corresponds to the critical point of the map: $z'(w)=0$. Then $w=a\pm
\sqrt A$ and $|w|=1$. Thus one can use eq.~(\ref{eq:1024}) to compute $z$
position of the cusp. It is given by
\begin{equation}
  \label{eq:53}
  z_{cr} = z(a\pm \sqrt A) = a\left( 1 + \frac A{a^2}\right)
\end{equation}
On the other hand we can rewrite position of the flux $\zeta$ as
\begin{equation}
  \label{eq:54}
   \zeta = a\left(\frac{1}{|a|^2}+\frac{A}{a^2(1-|a|^2)}\right)
\end{equation}
Due to comment after equation~(\ref{eq:51}) we see that equation in
parenthesis in both eqs.~(\ref{eq:53}) and~(\ref{eq:54}) are real. And thus,
position of the flux and cusp lie on the same ray, coming from origin.

The distance between the critical point and the position of the flux is always
positive.

\end{document}